\documentclass[preprint]{aastex}

\usepackage{amstext}
\usepackage{amssymb}
\usepackage{graphicx}
\usepackage{xspace}
\usepackage{enumitem}
\usepackage{xcolor}

\newcommand{\um}{$\mu$m\xspace}

\begin{document}

\title{The impact of cosmic rays on the sensitivity of JWST/NIRSpec}

\author{Giovanna Giardino$^1$, Stephan Birkmann$^2$, Massimo Robberto$^3$, Pierre Ferruit$^1$, Bernard J. Rauscher$^4$, Marco Sirianni$^2$, Catarina Alves de Oliveira$^2$,  Torsten Boeker$^2$, Nora Luetzgendorf$^2$, Maurice te Plate$^2$, Elena Puga$^2$, Tim Rawle$^2$} 
\affil{{\rm 1}. ESA, Science Operations Department, ESTEC, 2200AG Noordwijk, The Netherlands}

\affil{{\rm 2}. ESA, Science Operations Department, STScI, Baltimore, MD 21218, USA}
\affil{{\rm 3}. STScI, Baltimore, MD 21218, USA}
\affil{{\rm 4}. NASA Goddard Space Flight Center, MD 20771, USA}

\begin{abstract}


The focal plane of the NIRSpec instrument on board the James Webb Space Telescope (JWST) is equipped with two Teledyne H2RG near-IR
detectors, state-of-the-art HgCdTe sensors with excellent noise performance. Once JWST is in space, however, the noise level
in NIRSpec exposures will be affected by the cosmic ray (CR)
fluence at the JWST orbit and our ability to detect CR hits
and to mitigate their effect. We have simulated the effect of CRs on
NIRSpec detectors by injecting realistic CR events onto 
dark exposures that were recently acquired during the JWST cryo-vacuum
test campaign undertaken at Johnson Space Flight Center. Here we
present the method we have implemented to detect the hits in the
exposure integration cubes, to reject the affected data points within our
ramp-to-slope processing pipeline (the prototype of the NIRSpec official
pipeline), and assess the performance of this method for different
choices of the algorithm parameters.  Using the optimal
parameter set to reject CR hits from the data, we estimate
that, for an exposure length of 1,000\,s, 
the presence of CRs in space will lead to an increase of typically
$\sim 7\%$ in the detector noise level with respect to the on-ground
performance, and the corresponding decrease in the limiting sensitivity of the instrument, for the medium and high-spectral resolution modes.  

\end{abstract}


\section{Introduction}
\label{sec:intro}

The James Webb Space Telescope (JWST) is widely seen as the scientific successor to
the Hubble Space Telescope.  Optimized for near and mid-IR observations,
JWST is equipped with a 6.5\,m diameter primary mirror and is passively
cooled to less than 50~K \citep{gmc+09}. Scheduled for launch
in March 2021, the spacecraft will be placed in an orbit around the
Sun--Earth Lagrange point L2. The JWST project is led by
the National Aeronautics and Space Administration (NASA), with major
contributions from the European Space Agency (ESA) and the Canadian Space Agency (CSA).  The observatory will carry a suite of four science instruments, one of
which is the Near Infrared Spectrograph (NIRSpec), developed by ESA
with Airbus Defence and Space Germany as the prime contractor
\citep{bkf+2007, bfr+2016}. The primary goal of NIRSpec is to enable
large spectroscopic surveys in the near-infrared with an emphasis on
the study of the birth and assembly of galaxies.

To detect and characterize extremely faint astronomical objects, such as
primordial galaxies, NIRSpec has to achieve very high sensitivity and
thus its detector noise performance is crucial. For this
reason, the NIRSpec focal plane is equipped with ultra low noise
near-IR detectors: two Teledyne H2RG 2048$\times$2048
pixel, 5.3 \um-cutoff detectors, provided by NASA's Goddard Space Flight Center (GSFC)
\citep{beletic+08, Rauscher+2014}. The other two near-infrared science instruments 
onboard JWST (NIRCam and NIRISS) have similar detectors. The future ESA missions Euclid and Ariel will also be equipped with similar detectors and the next generation of the Teledyne HxRG family (H4RG) are planned for NASA's Wide Field Infrared Survey Telescope (WFIRST).

As is often the case for IR detectors, both NIRSpec detectors (NRS1 and
NRS2 hereafter), are read out non-destructively or ``up-the-ramp''. An exposure comprises one or more ramps, also referred to as ``integrations". An
integration consists of one or more “groups” (including the initial
reset), and each group consists of one frame or the on-board average of
multiple frames -- for details, see \cite{rff07, rff10err}.  For the full-frame mode
that will be used to study faint astronomical objects, NIRSpec offers two
readout options: the so-called traditional mode and the IRS$^2$ (Improved Reference Sampling and Subtraction) readout mode, developed by the NASA/GSFC team to minimize the presence of correlated noise caused by the detectors readout electronics
\citep{raf+2017}. For each readout mode, two sampling patterns (with and without frame-averaging data) are available: NRS (averaging 4 frames per
group) and NRSRAPID (1 frame per group) for traditional mode, and
NRSIRS2 (averaging 5 frames per group) and NRSIRS2RAPID (1 frame per group) for
IRS$^2$ mode -- see also the JWST User
Documentation\footnote{https://jwst-docs.stsci.edu}.

\cite{bsf+2018} have recently re-assessed the noise performance of
NIRSpec detectors, using data acquired during cryogenic testing of the JWST OTIS\footnote{OTIS
  denotes the JWST element comprising the Optical Telescope Element
  (OTE) and the Integrated Scientific Instrument System (ISIM). This
  is the part of the observatory that will lie in the permanent shadow of the JWST sun-shield,
  and hence will be passively cooled to $\sim$50 K} element at Johnson Space Center in Summer 2017: 25 dark exposures in NRSRAPID mode with one integration
of 88 groups each, and 30 exposures in NRSIRS2RAPID mode with one
integration of 200 groups each. They conclude
that the NIRSpec detector system meets its stringent noise requirement
of 6 electrons total noise in a $\sim$1,000 s exposure.  The question,
however, is whether this performance will also be achieved once the instrument is in
space and subjected to the intense CR flux expected for the L2 orbit. In this paper, we will address what type of performance degradation is to be expected due to the CR environment, and whether we have the tools to effectively minimize its detrimental effect on the NIRSpec performance.

In order to quantify the impact of CRs on NIRSpec
observations, we have adopted a library of simulated CR
events on IR detectors which was originally developed to test and optimize the JWST data processing pipeline
\citep{Robberto2010}. We used the library to populate with CR hits the NIRSpec dark
exposure data acquired during the OTIS test campaign. We then
implemented a method to identify the hits in the integration ramps and
remove them from the data cubes within the 'ramp-to-slopes' part of
the processing pipeline developed by the ESA NIRSpec team (which
serves as template for the official NIRSpec pipeline developed by
STScI). Our detection algorithm is based on the two-point difference method \citep{osf+1999, foh+00} and relies on thresholds tuned to achieve the best trade-off in terms of residual events and loss of effective
integration time. We  quantify the impact on the
noise performance of the NIRSpec detectors from the CR bombardment expected for the JWST orbit by comparing the processed (CR-injected and removed) data with the original data.

The paper is organized as follows: the simulations of the CR events are
presented in Sect.\,\ref{sec:sim}, while the CR detection and
rejection algorithm is described in Sect.\,\ref{sec:algo}; the results
in terms of residual noise levels for the different readout modes are
summarized in Sect.\,\ref{sec:res} and discussed in
Sect.\,\ref{sec:disc}.

\section{Data and Simulated Cosmic Ray Events}
\label{sec:sim}

\subsection{Cosmic Ray library}
\label{sec:crlib}
\cite{Robberto2010} presents a library  of simulated cosmic ray events on JWST HgCdTe detectors. 
The cosmic rays are calculated for three different levels of solar activity, namely a) Solar minimum and Galactic maximum, b) Solar maximum and Galactic minimum, and c) Solar Flare, corresponding to
three main space ionizing-radiation environments predicted by the CREME96 model \citep{CREME96}. The calculation takes into account the relative frequency of the most abundant nucleons: H, He, C, N, O, Fe, and their energy. Assuming a shielding of 100mil Al equivalent thickness, the model predicts a flux of 4.9, 1.8, and 3046 events~${\rm cm^{-2}~s^{-1}}$ respectively for the three levels of solar activity. The flux at Solar minimum is consistent with a typical particle background at L2 of 5 protons~${\rm cm^{-2}~s^{-1}}$ reported by the Planck and Gaia missions \citep{caa+2014, ckh+2016}.

\begin{figure}[t]
\centering
\includegraphics[width=0.7\textwidth]{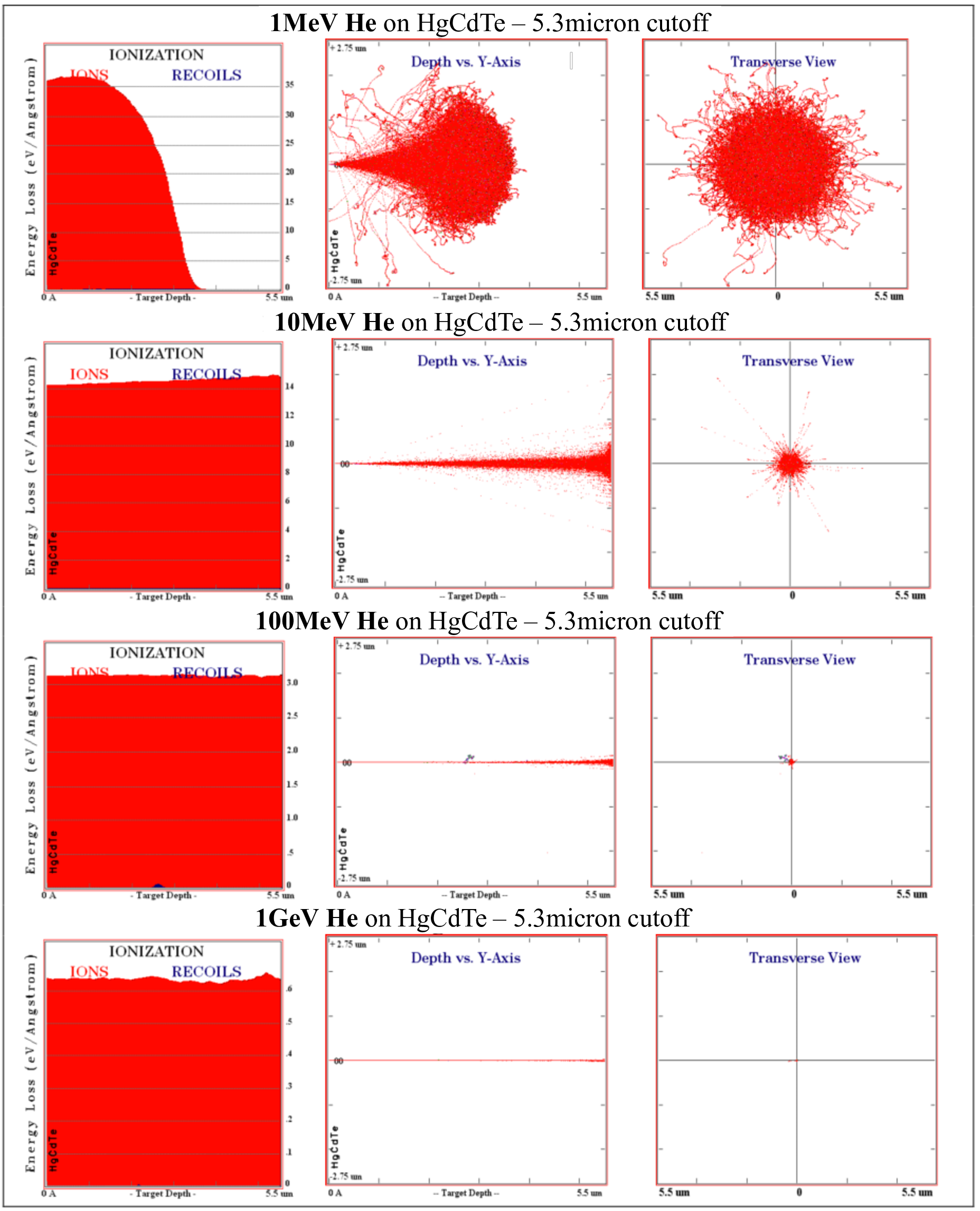}
\caption{\label{fig:He_on_MCT5}TRIM Results for He nuclei of different energy (from top to bottom: 1MeV, 10MeV, 100MeV, 1GeV) impacting on 5.3 $\mu{\rm m}$ HgCdTe. The left column shows the energy loss, the central column shows a longitudinal view with generated charges, the right column shows the transverse view of the generated charges.}
\end{figure}

Cosmic rays impacting the detectors lose energy mostly through inelastic collisions with the bound electrons of the detector material. The detailed calculation of the energy loss is performed using SRIM \citep[Stopping and Range of Ions in Matter,][]{2010NIMPB.268.1818Z} for different types of JWST detectors, characterized by the stoichiometric ratio of Hg vs. Cd (controlling the band gap of the HgCdTe alloy and therefore the long wavelength cutoff of the material), the material density (depending on the stoichiometric ratio) and thickness, assumed to be ``optimally'' tuned to be equal to the long wavelength cutoff of the material (5.3 $\mu$m in the case of the NIRSPEC detectors). The module TRIM (Transport of Ions in Matter) of the SRIM package allows the calculation of the 3D distribution of the ions in the material, together with the kinetic phenomena associated with the ion's energy loss: target damage, sputtering, photon production and ionization, this last effect being the one of interest.
Figure \ref{fig:He_on_MCT5} shows an example of the TRIM calculations; the penetration of CRs into the material depends on their energy. The fractional energy loss is higher at low energy, as predicted by the classic Bethe-Bloch formula. 

The CR library is built performing Monte Carlo simulations of CRs generated by the CREME96 model that  impact a random point within a pixel with a spatially isotropic distribution. The energy losses are determined following the CR through the pixels that are geometrically intersected, assuming  to have a square 18~$\mu$m$\times18~\mu$m surface and 5.3~$\mu$m thickness; grazing CRs can affect two or more pixels.
For each event, the simulation calculates the electrons generated in a $21\times21$ pixel region centered around the point of impact at pixel (10,10); this area is large enough to account for nearly tangential events and Inter Pixel Capacitance (IPC) coupling, i.e. the capacitive coupling arising between neighboring detector pixels in the source-follower CMOS design,  via displacement currents flowing from the collection node \citep{mnf+04}. 

\begin{figure}[t]
\centering
\includegraphics[width=0.7\textwidth]{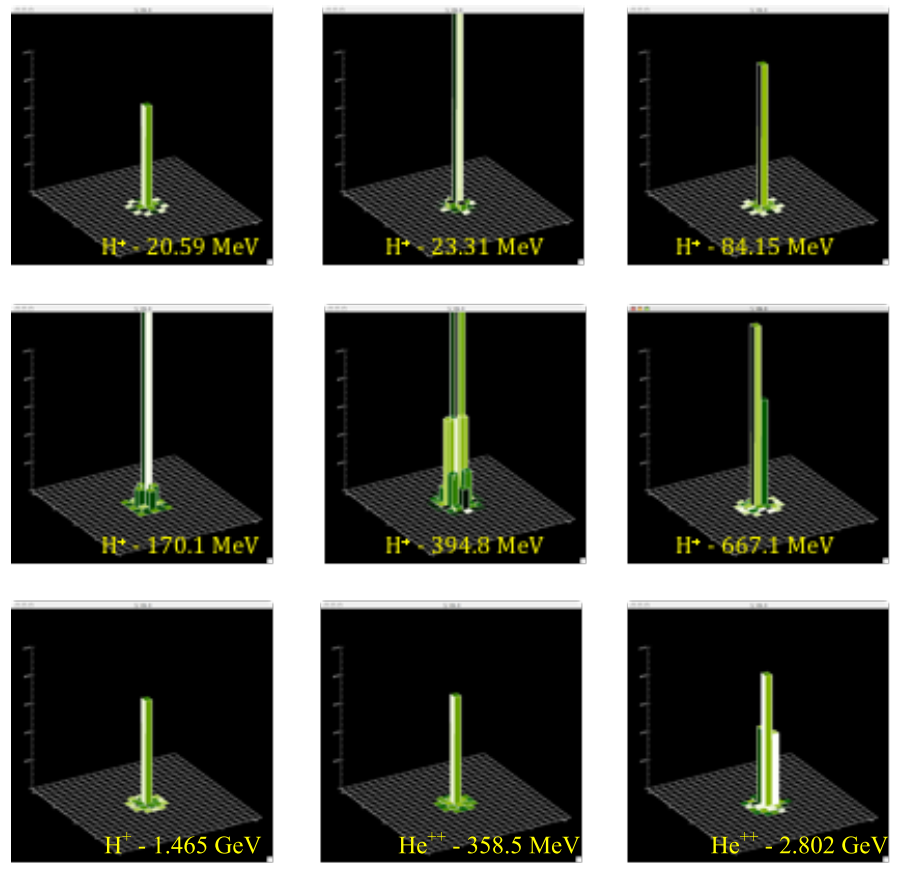}
\caption{\label{fig:CRs_on_MCT}Simulated CR events on HgCdTe detectors for different energy levels of H and He nucleons, and including capacitive coupling.}
\end{figure}

The effect of IPC was taken into account by convolving each CR original image  with a 5x5-IPC kernel $F(i,j) = \alpha^{d(i,j)} $,
where $d$ is the distance from the central pixel and $F(0,0)$ is normalized  to preserve the flux. The measured value of the coupling factor for our detectors varies from 0.003 to 0.007 across the four outputs. To be conservative, we assumed $\alpha = 0.007$, resulting in the kernel shown in Table\,\ref{tab:kernel}: 

\begin{table}[h]
\centering
\footnotesize{
\begin{tabular}{|c|c|c|c|c|}
\hline
 8.036E-07 & 1.519E-05 &  4.9E-05 & 1.519E-05 & 8.036E-07\\
\hline
 1.519E-05 & 8.964E-04 &    0.007 &  8.964E-04 & 1.519E-05\\
\hline
 4.9E-05 &    0.007 &   0.9681 &    0.007 &  4.9E-05\\
\hline
1.519E-05 & 8.964E-04 &    0.007 & 8.964E-04 & 1.519E-05\\
\hline
8.036E-07 & 1.519E-05 &  4.9E-05 & 1.519E-05 & 8.036E-07\\
\hline
\end{tabular}
}
\caption{\label{tab:kernel}
5$\times$5-Kernel used to convolve the simulated CR-images and mimic the detector IPC}
\end{table}

The library contains 30,000 CR events in the form of 21x21 postage-stamp arrays. In Figure~\ref{fig:CRs_on_MCT} we present a few examples. For the analysis presented in this paper, we used the
fluence level at Solar minimum and Galactic maximum, which is higher than the one for Solar maximum and Galactic minimum and therefore represents a conservative assumption for the typical (non-Solar flare) operating conditions of JWST.

\subsection{Data processing}
\label{sec:proc}

As mentioned before, we used 25 NRSRAPID dark exposures with 88 groups each (corresponding to $\sim$ 945\,s integration time) and 30 NRSIRS2RAPID dark exposures of 200 groups each ($\sim$ 2,903\,s integration time), 
acquired during OTIS testing at Johnson Space Center in Summer 2017.

CR events were injected into the dark exposures using 
scripts operating on the individual frames of an exposure ramp. The number of injected CR events per frame is determined by the integrated fluence (4.9 events~${\rm cm^{-2}~s^{-1}}$ assuming 100 mil Al equivalent thickness shielding), multiplied by the H2RG detector area, (i.e. 13.5895 cm$^{2}$) and the frame readout time ($\sim$10.7 s for traditional readout and $\sim$14.6 s for IRS$^2$), resulting in $\sim$ 710
and 970 CR hits per frame in Traditional and IRS$^2$ readout modes respectively. 
For each frame, the events are randomly drawn from the library and
randomly placed (with a uniform distribution) across the detector arrays. The $21\times21$ pixel large electron-rate 'postage stamp' associated with each CR is converted into counts (using each detector's average gain value) and added to the frame, resulting in a 'jump' in the ramp of the affected pixels. If the maximum-value (i.e. saturation value) of the 16-bit analog-to-digital converter is reached, the ramp value is set to 65,535 from that sampling point onward.  The CR-count injection onto the frame is a simple addition, with no attempt to account for the  non-linear pixel response, which typically matters for count levels above $\sim$~50,000 counts, or other effects such as, e.g. persistence (see also the discussion in Sect.\,\ref{sec:disc}). CRs
were only injected into the light sensitive pixels (i.e. the central 2040 x 2040
pixel region of each detector), but not the 4-pixel wide 'frame' of reference pixels. 

\begin{figure}[t]
\centering
\includegraphics[width=0.48\textwidth]{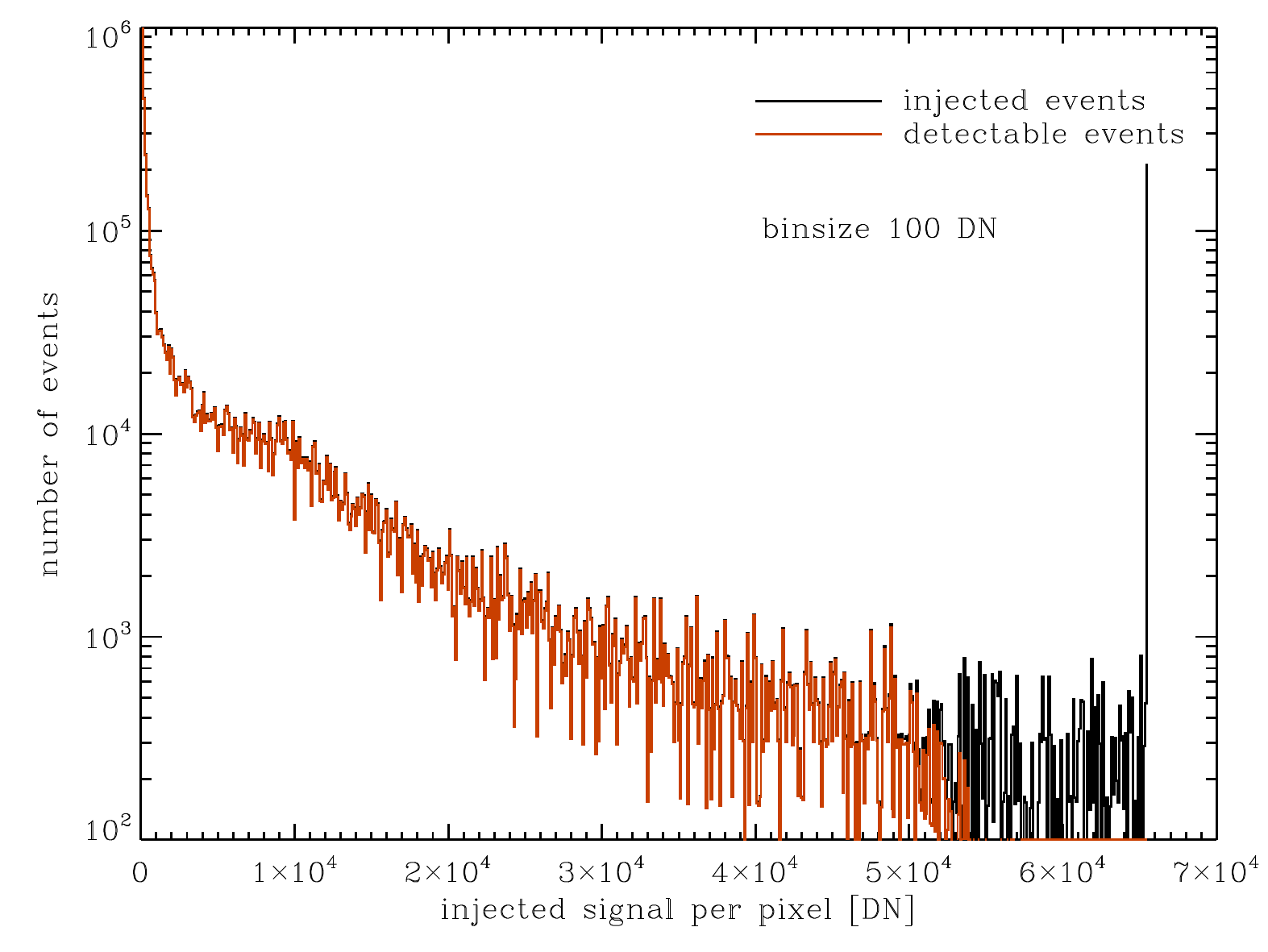}
\includegraphics[width=0.48\textwidth]{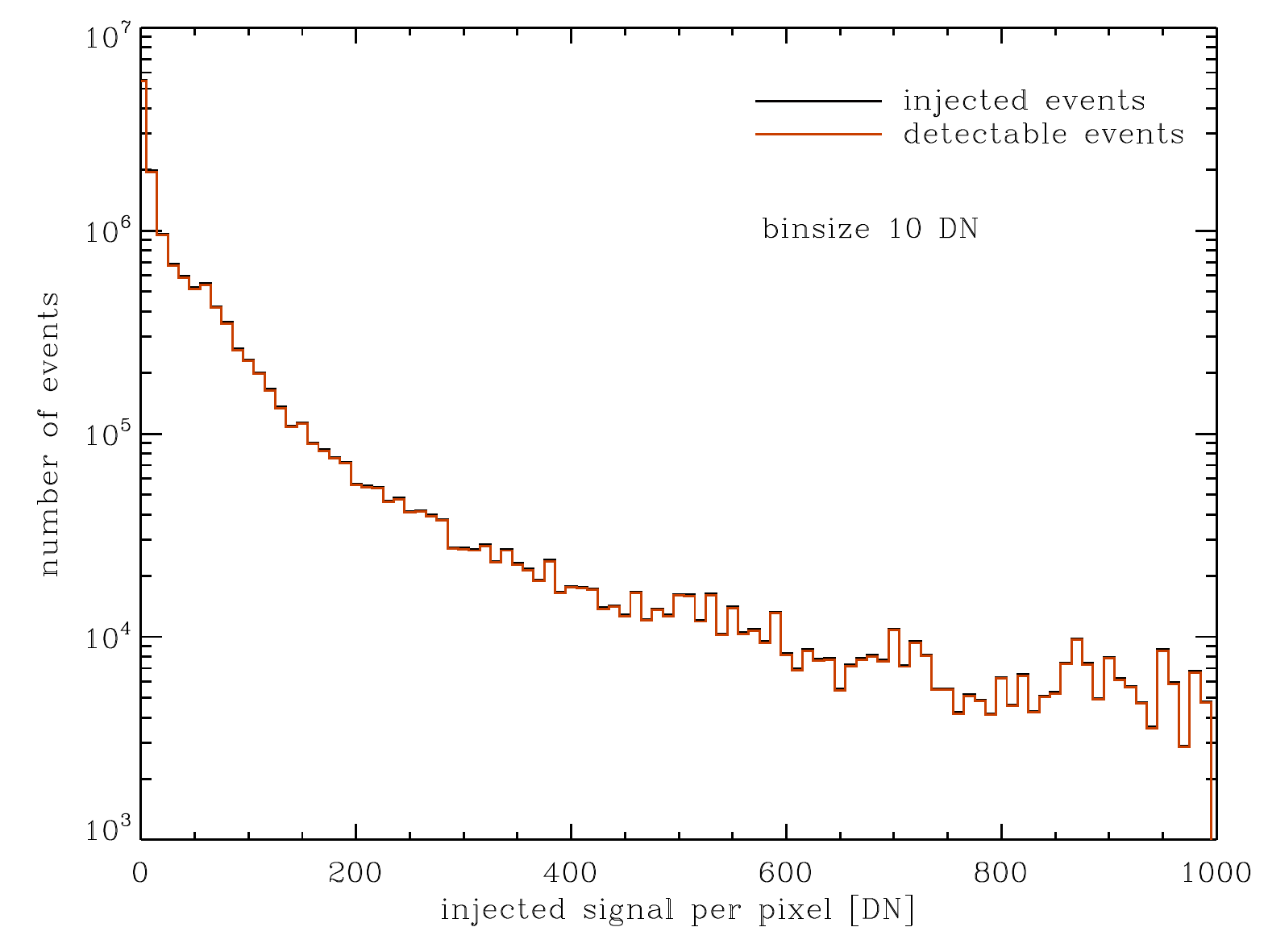}
\caption{\label{fig:hist}Histogram of the injected events for the NRS1
  exposures. Complete histogram with 100 DN bin size is on the left, a
  more detailed view for the low levels at the right. The black curve
  corresponds to all injected events, whereas the red curve is for
  detectable events only. Events above $\sim$55,000 counts can never
  be detected by the pipeline, because when added to the bias level
  they exceed the range of the analog to digital converter, and are
  therefore flagged as saturated.}
\end{figure}

Throughout the duration of an exposure, the number of pixels affected by a CR hit increases with the integration time: at the end of a traditional-mode 88-groups
integration, $\sim$ 15\% of pixels have been affected by at least one
hit, while at the end of a IRS$^2$-mode 200-groups integration $\sim$ 40\% of
pixels have been affected. Fig.\,\ref{fig:hist} shows the
histogram of the injected events for the dark exposures of NRS1. The
distribution of charge injected into individual pixels steeply
increases at low count levels. The event histograms for NRS2
look very similar, and are not shown.

All the exposures were processed with the ESA NIRSpec ramp-to-slope
pipeline that performs the following basic data reduction steps.
\begin{itemize}
\item{{\em Saturation detection and flagging}: each pixel value in each group
is compared to the pixels' saturation threshold stored in a reference
file, and flagged as saturated if it exceeds the threshold. Data
flagged as saturated are not used when determining the count
rate.}
\item{{\em Master bias subtraction}: the master bias, i.e. the detector frame readout at zero seconds, derived from averaging the first frame of many dark exposures, is subtracted
from each group, pixel-by-pixel.}
\item{{\em Reference pixel subtraction}: for Traditional readout, the
four rows of reference pixels at the top and bottom of the detector
(eight rows in total) are used to determine the pedestal for each
group and output. The average pedestal is subtracted from all pixels
in a given output in each group. Finally, a correction along the slow
readout direction is performed using the four columns of reference
pixels to the left and right and a running windowed average. For IRS$^2$
readout, the interleaved reference pixels and reference output are subtracted from all pixels using a statistically optimized set of frequency dependent weights to better correct for $1/f$-noise and correlation in the detector data, as described in \citet{raf+2017}.}
\item{{\em Dark subtraction:} the reference dark cube is
subtracted from each group, pixel-by-pixel.}
\item{{\em Count rate estimation, including jump detection and CR rejection:} in
this step, the count rate per pixel is estimated using all unsaturated
groups in the ramp, and 'jumps' (e.g.\ due to CR hits) are
detected and rejected by estimating the slope of individual ramp segments on either 
side of the affected frame. More details on this step are
provided in the next section.}
\end{itemize}

\section{Detection and Rejection of CR events}
\label{sec:algo}

To illustrate the slope estimation algorithm used by the ESA pipeline,
we assume an integration with $n$ unsaturated groups with $m$ frames
each (for NIRSpec $m = 1, 4,$ or $5$, depending on the readout mode
and pattern). For a given pixel, the measured signal at each
group $i = 1\ldots n$ is $s_i$ (in DN, i.e. Digital Number\footnote{also referred to as ADU, i.e. Analog-to-Digital Unit}). The count rate $b$ in electrons per
second for that pixel is then
\begin{equation} \label{eq:slope}
b = \frac{g}{t_g}  \sum_{i=1}^n a_i s_i ,
\end{equation}
where $g$ is the conversion gain in electrons/DN for the given pixel,
$t_g$ is the group time in seconds, and $a_i$ are pre-computed and
normalized weights (their sum normalized to 1). The weights depend on
the read noise $\sigma$, the estimated signal $s$, and the number of
groups in an integration $n$. Details on how the weights are
calculated are given in \cite{foh+00}, who show that optimally
weighting is needed to efficiently deal with different regimes of
signal-to-noise.  In the low signal to noise limit where $b\times t_g
\ll \sigma^2$ each group is weighted the same. This is equivalent to
uniform weighting and a linear fit to the data. In the high signal to
noise regime where $b\times t_g \gg \sigma^2$ the full weight is on
the endpoints $s_1$ and $s_n$ and the data in between are basically
ignored, optimizing effective integration time -- see also
\cite{Robberto2014} on the issue of generalized least square in
non-destructive readouts.

CR hits on the pixel during the integration will deposit
additional charge which will lead to a jump or discontinuity in the
ramp and skew the estimated slope if not corrected for. If an
integration has enough groups ($n \geq 3$), one can evaluate the
$(n-1)$ pair-wise differences (or correlated double sample, CDS) in
order to look for outliers:
\begin{equation} \label{eq:diff}
d_{i=1\ldots n-1} = s_{i+1} - s_i.
\end{equation}
If only two data points are available ($n=2$), no outliers detection
can be performed as there is only one pair, and the signal is directly
calculated using equation~\ref{eq:slope}.  A first robust estimate of the average difference /
signal between groups is obtained by averaging the differences after removing the
most extreme values:
\begin{equation}
d_{avg} = \left\{
\begin{array}{ll}
\frac{1}{n-2}\left((\sum_{i=1}^{n-1} d_i) - d_{max}\right)&\mathrm{for}\; n = 3\\
\frac{1}{n-3}\left((\sum_{i=1}^{n-1} d_i) - (d_{max} + d_{min})\right)&\mathrm{for}\; n > 3\\
\end{array}
\right\},
\end{equation}
where $d_{max}$ and $d_{min}$ are the maximum and minimum of all the
differences $d_i$, respectively. This is done to remove the
contribution of potential outliers from the average. The expected
variance of the two-point differences $d_i$ is then
\begin{equation}
\sigma_{cds}^2 = \frac{2}{m}\sigma_{read}^2 + g\left(1-\frac{m^2 - 1}{3m^2}\right)d_{avg}.
\end{equation}

In order to detect significant outliers from this, we calculate the
deviations from the mean difference and the maximum absolute outlier
\begin{equation}
\delta_j = max(|\delta_{i=1\ldots n-1}|) = max(|d_{i=1\ldots n-1} - d_{avg}|)
\end{equation}
and store the position of the maximum absolute difference as
$j$. $\delta_j$ is then compared to the expected noise times a
threshold scaling factor $c_{th}$ and flagged as an outlier if the
noise threshold is met or exceeded:
\begin{equation}\label{eq:thres}
| \delta_j | \ge  c_{th} \sigma_{cds}.
\end{equation} 
If the worst offender exceeds the set threshold, the integration will
be split into two segments at difference $j$, and the count rate $b_1$
and $b_2$ of those segments will be determined independently:
\begin{equation}
\begin{array}{ll}
b_1&= \frac{g}{t_g}  \sum_{i=1}^j a_i s_i ,\\
b_2&=\frac{g}{t_g}  \sum_{i=j+1}^n a_i s_i .\\
\end{array}
\end{equation}
The ramp segments have $n_1 = j$ and $n_2 = n - j$ data points,
respectively. These numbers are used to determine the optimal weights
$w_1$ and $w_2$ for combining the ramp segments count rates according
to \cite{osf+1999}:
\begin{equation}
\begin{array}{ll}
w_1&=  (n_1 - 1)n_1(n_1 + 1) ,\\
w_2 &= (n_2 - 1)n_2(n_2 + 1) ,\\
\end{array}
\end{equation}
with the final / combined count rate being $b_f$:
\begin{equation} \label{eq:comb}
b_f = \frac{w_1 b_1 + w_2 b_2}{w_1 + w_2} .
\end{equation}

The above approach effectively mitigates one significant
(i.e.\ detected) outlier in an integration. To be able to
detect multiple events, the above (equations~\ref{eq:diff} through
\ref{eq:comb}) are recursively repeated for the ramp segments $i =
1\ldots j$ and $i = j+1 \ldots n$, so that outliers can be detected in
these sub-ramps as well. This recursive process is stopped when either
no outliers are detected in the sub-ramp anymore or the number of
groups per (sub-)ramp becomes $n \leq 2$.
The points in time (positions in the ramp) where these outliers are
detected can be stored to be able to
compare the masks of simulated injected CR events with the ones
detected by the pipeline.

The process outlined above is performed for each pixel in the detector
individually. However, it is expected that typical CR events
will affect a group of adjacent pixels rather than a single one mainly
because of the coupling due the IPC, but also because of the
intrinsic shape of the CR footprint. Therefore, the pipeline offers
the possibility to identify neighbors of (clusters of) pixels and
perform a second detection and rejection cycle with a different
threshold on those pixels.

We investigated two options for selecting neighbors: {\em i)} directly
adjacent pixels of any pixel that has a CR event of at least a
certain magnitude, or {\em ii)} neighbors adjacent to a cluster of
pixels with detections at the same group. A cluster consists of two or
more pixels that are directly adjacent, i.e. neighbor at the top,
bottom, left or right. Pixels that are located diagonally across are not considered 
neighbors nor members of the same cluster, as illustrated
in Fig.\,~\ref{fig:clusters}. 

It is expected that most CR events of significant magnitude will
impact more than one pixel, therefore only considering neighbors
around clusters of two or more adjacent pixels for a second pass might
be sufficient. Also, both NIRSpec detectors have a small population of
pixels exhibiting Random Telegraph Noise (RTN), a phenomenon in which a pixel counts transition rapidly, and almost `digitally', between two levels \citep{bmp+2004, rfr+2004}. Our approach of flagging only cluster neighbors therefore has the advantage of
limiting the impact of RTN pixels affected on their neighbors, i.e.\ not unnecessarily breaking up ramps. On the other hand, the cluster approach potentially leads to issues caused by the effects of IPC, because the neighbors of single pixel events will not be flagged, regardless of how strong the event was. Therefore, the pipeline also
allows the flagging of neighbors that are next to individual pixels
that have been flagged, as long as the jump in the latter is of
sufficient magnitude. The default jump threshold was selected to be 200 DN, which is large enough to exclude most RTN pixels.

The second detection pass on single-pixel neighbors is then performed
with a threshold $c_{nb}$ using equation~\ref{eq:thres}, with $c_{nb}$
replacing $c_{th}$, and only considering the two-point differences
where the parent CR of the adjacent cluster/pixel took place
(there can be more than one event in a ramp). To have any effect $c_{nb}$ 
has to be lower than $c_{th}$  (otherwise the event
would have been found in the first pass already) and it can be as low as
zero, which means that the ramps of the neighbors will always be
broken into sub-ramps at the position of the parent event (i.e. the
condition in equation~\ref{eq:thres} is always fulfilled). The two or
more sub-ramps are then individually processed and combined to a final
count rate using the weighting scheme outlined in
Sect.~\ref{sec:proc}.

We evaluated the efficiency of detecting CR hits for different values of
$c_{th}$ and $c_{nb}$ as described in the Appendix. 

\section{Results\label{sec:results}}
\label{sec:res}

\subsection{Traditional readout}
\label{sec:trad}

\begin{figure}
\centering
\includegraphics[width=0.48\textwidth]{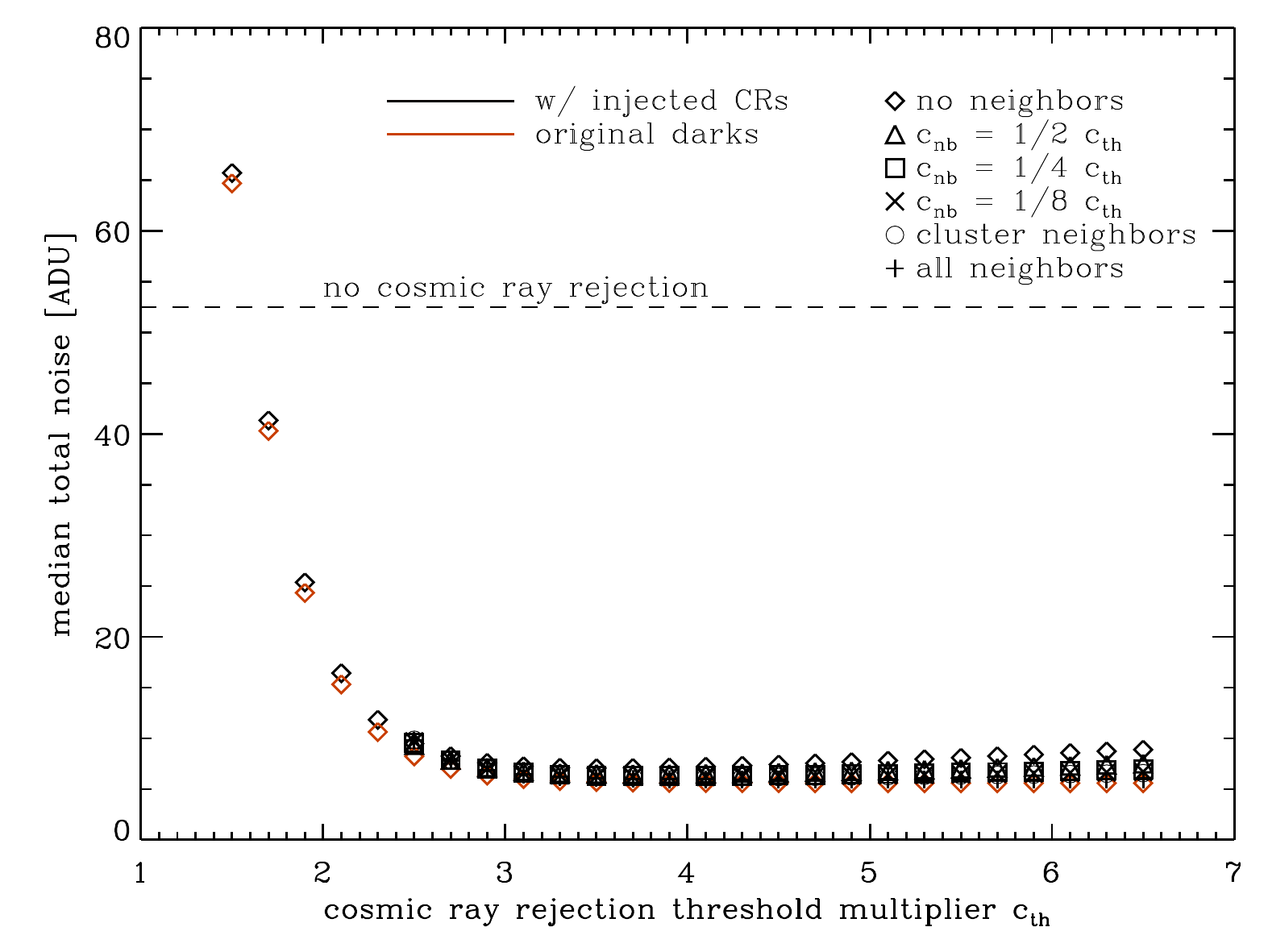}
\includegraphics[width=0.48\textwidth]{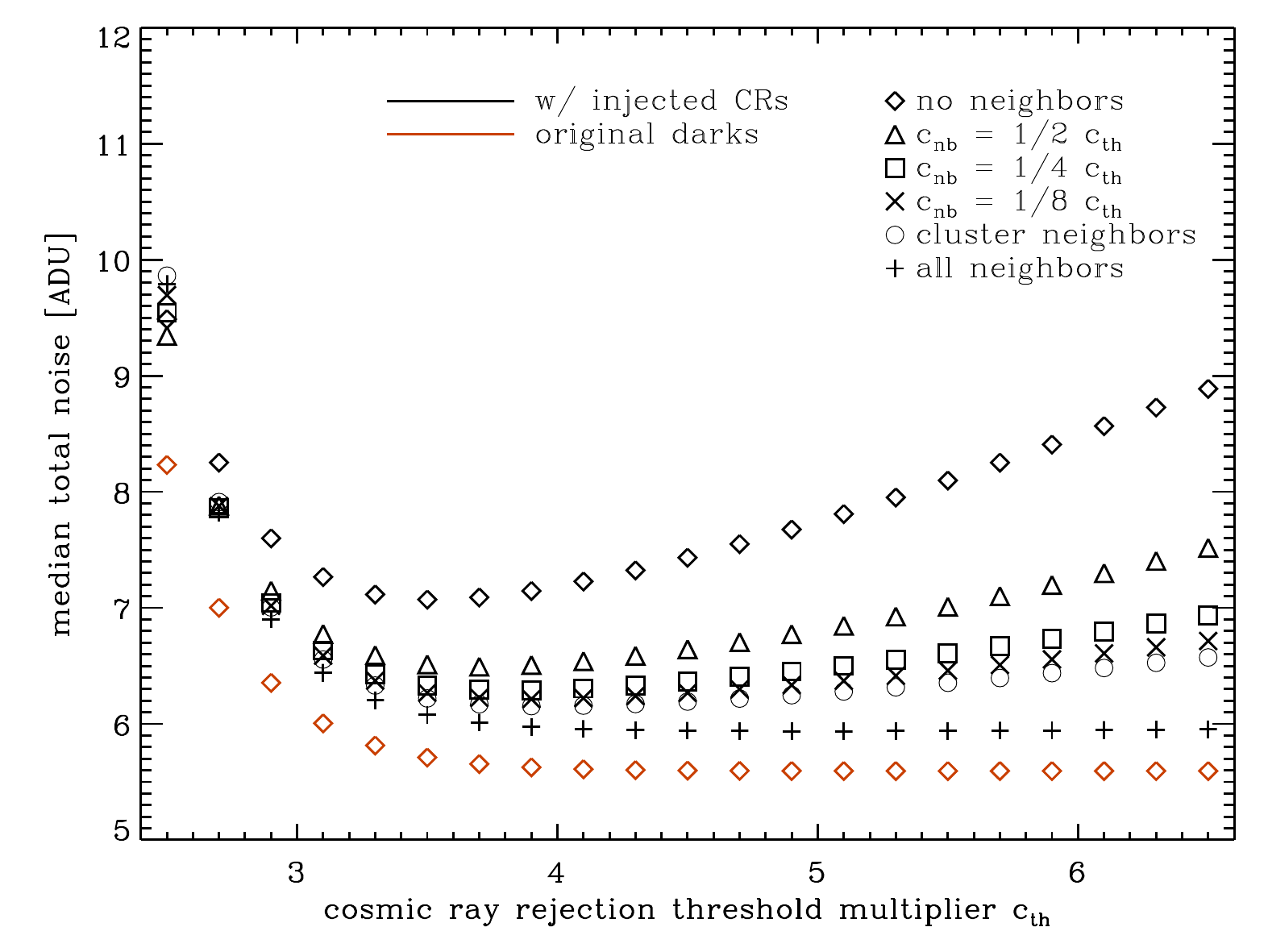}
\caption{\label{fig:noise1}
Median total noise for NRS1 for the 25 dark exposures with (black
symbols) and without (red symbols) simulated CR events. The
dashed black line in the left panel shows the noise with no CR
rejection at all ($c_{th} = \infty$). The right panel gives a more
detailed view of the lower total noise range.}
\end{figure}

\begin{figure}
\centering
\includegraphics[width=0.48\textwidth]{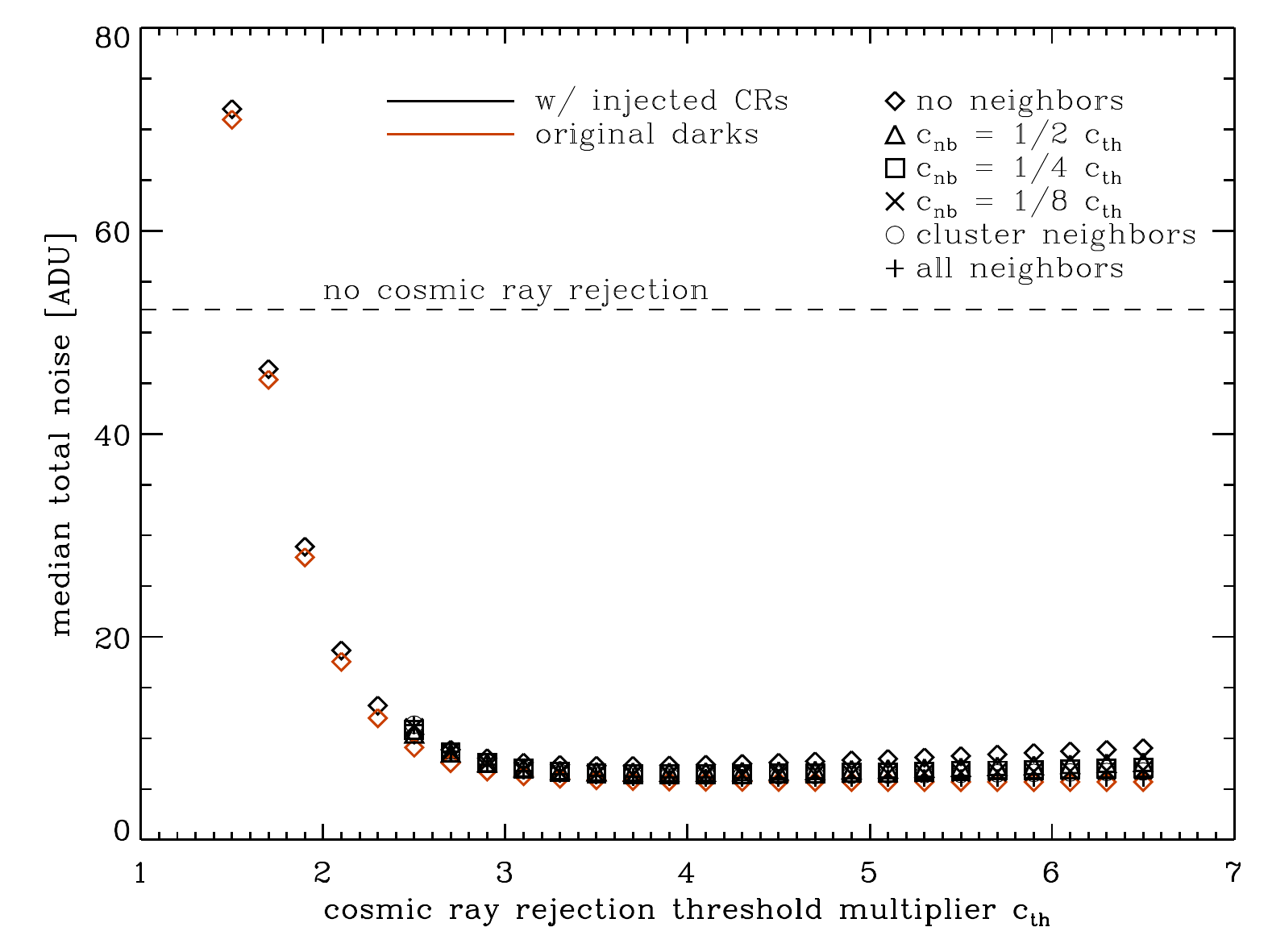}
\includegraphics[width=0.48\textwidth]{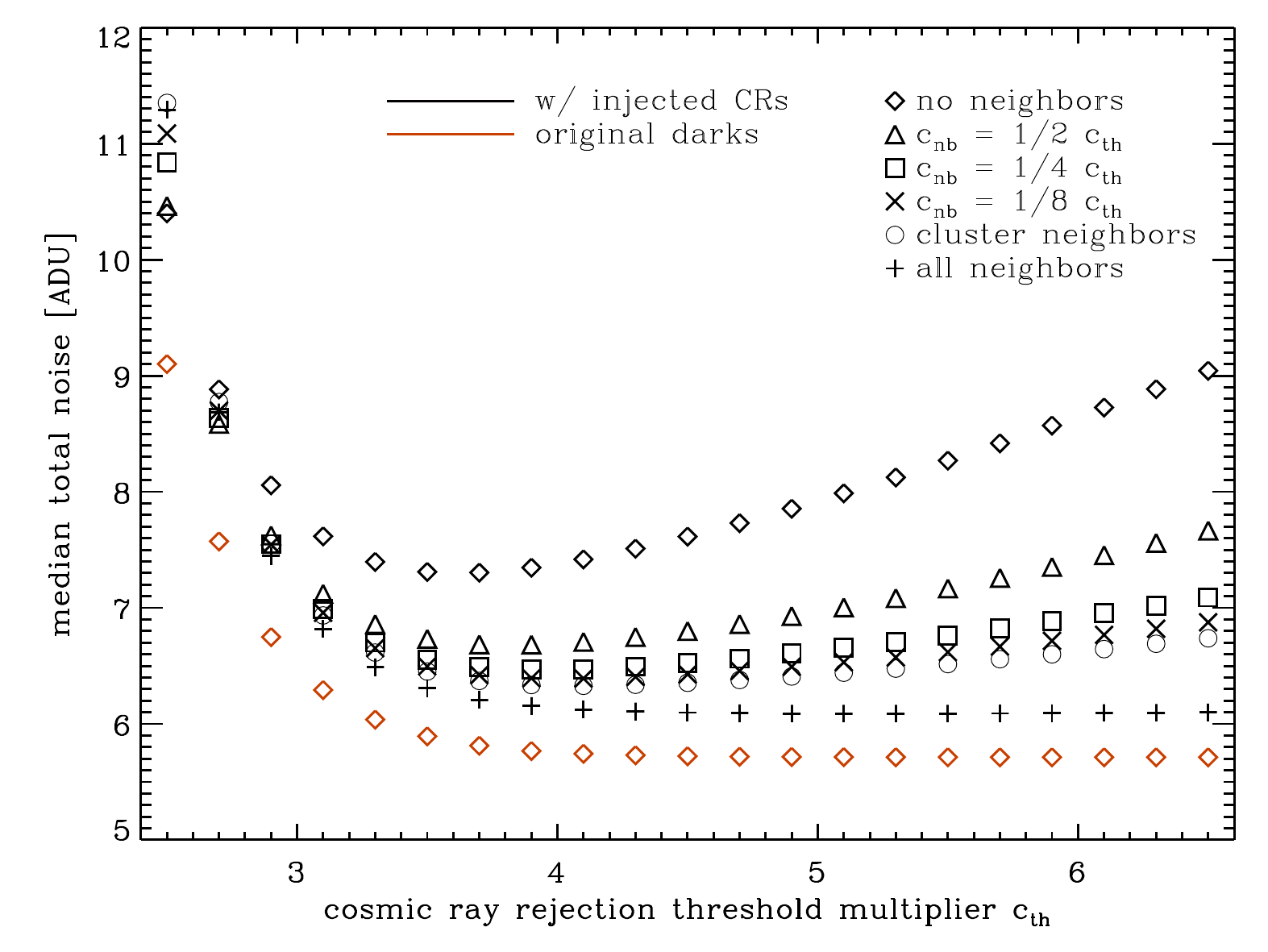}
\caption{\label{fig:noise2}
Median total noise for NRS2 for the 25 dark exposures with (black
symbols) and without (red symbols) simulated CR events. The
dashed black line in the right panel shows the noise with no CR
rejection at all ($c_{th} = \infty$). The right panel gives a more
detailed view of the lower total noise range.}
\end{figure}

In order to evaluate the impact of CR hits on the NIRSpec sensitivity, we evaluated the typical noise in dark exposures by first calculating the standard deviation per
pixel across all 25 processed exposures, and then deriving the median value across all $2040^2$ pixels. This was done for both the original darks and the CR-injected darks, the latter being processed for different combinations of $c_{th}$ and $c_{nb}$. 
The total noise in e$^-$ was then
calculated by multiplying the median standard deviation with the
effective integration time $t_{eff} = \left( n - 1 \right) t_g$ and the conversion gain (e$^-$/DN).
The results are presented in Figures~\ref{fig:noise1} and
\ref{fig:noise2}.  The lowest total noise is achieved for threshold
multiplier values of $c_{th} \gtrsim 4$ for all neighbor detection
thresholds. The total noise degrades rapidly with lower
thresholds. Higher thresholds have less impact on the achieved noise,
and are in fact preferred for the original data without the added
simulated CR events, although the total noise curve is basically flat
for $c_{th} > 4$.

The reason for the observed behavior can be interpreted as follows:
as apparent from the plots in Appendix~\ref{sec:efficiency}, a low detection
threshold $c_{th}$ will lead to many ``false positive''
detections. Due to the symmetric nature (both positive and negative
outliers are flagged) of the detection algorithm, this should not
introduce a bias to the estimated slope. However, it will have a
negative effect on the achieved signal-to-noise, because integrations
are unnecessarily broken up into many shorter sub-ramps, which is
particularly detrimental in the low signal-to-noise regime (e.g.\
darks), where the read noise dominates. At higher thresholds, a
smaller and smaller fraction of affected pixels is actually detected,
leading eventually to an increase in total noise.

It is also evident that enabling a secondary pass on neighbors of
detected pixel clusters with a lower threshold reduces the total
noise, as long as the initial threshold $c_{th}$ is not too low. In
Table~\ref{tab:res} we list the best parameter values (i.e. yielding
the lowest noise) for each tested method of neighbor detection in the
two detectors.

\begin{table}[h]
\centering
\begin{tabular}{|l|r|r|r|r|}\hline
\multicolumn{1}{|c|}{$c_{nb}$}&\multicolumn{2}{c|}{NRS1}&\multicolumn{2}{c|}{NRS2}\\
&$c_{th}$&noise [e$^-$]&$c_{th}$&noise [e$^-$]\\\hline\hline
$\infty$ (no neighbors)&3.5&7.038&3.7&8.294\\
$1/2 \times c_{th}$&3.7&6.463&3.9&7.589\\
$1/4 \times c_{th}$&3.9&6.261&3.9&7.343\\
$1/8 \times c_{th}$&3.9&6.184&4.1&7.258\\
$0$ (all cluster neighbors)&3.9&6.125&4.1&7.186\\
$0$ (all neighbors)&4.9&5.910&5.1&6.914\\\hline
original data (no neighbors)&5.3&5.568&5.7&6.487\\\hline
\end{tabular}
\caption{\label{tab:res}
Best parameters for CR detection threshold multiplier $c_{th}$
and the associated total noise for different values of $c_{nb}$ for
the two NIRSpec detectors. The last row gives the total noise measured
for the original dark data without simulated CRs added.}
\end{table}

The lowest total noise is achieved using a second CR rejection iteration
on all neighbors (i.e.\ with a threshold of $c_{nb} = 0$), including
neighbors of single pixel events that are marked as severe. That means
that for these neighbors, the ramps are broken at the same position as
the parent cluster/pixel, regardless of the two-point
difference. Given the steep rise of total noise towards lower values
of $c_{th}$ and the little impact of using higher and higher threshold
values, in particular when flagging all neighbors (see
Figures~\ref{fig:noise1} and \ref{fig:noise2}), the best combination
of threshold values are $c_{th} = 5.0$ and $c_{nb} = 0$ with single
pixel neighbor flagging, for both detectors.

\begin{figure}
\centering
\includegraphics[width=0.48\textwidth]{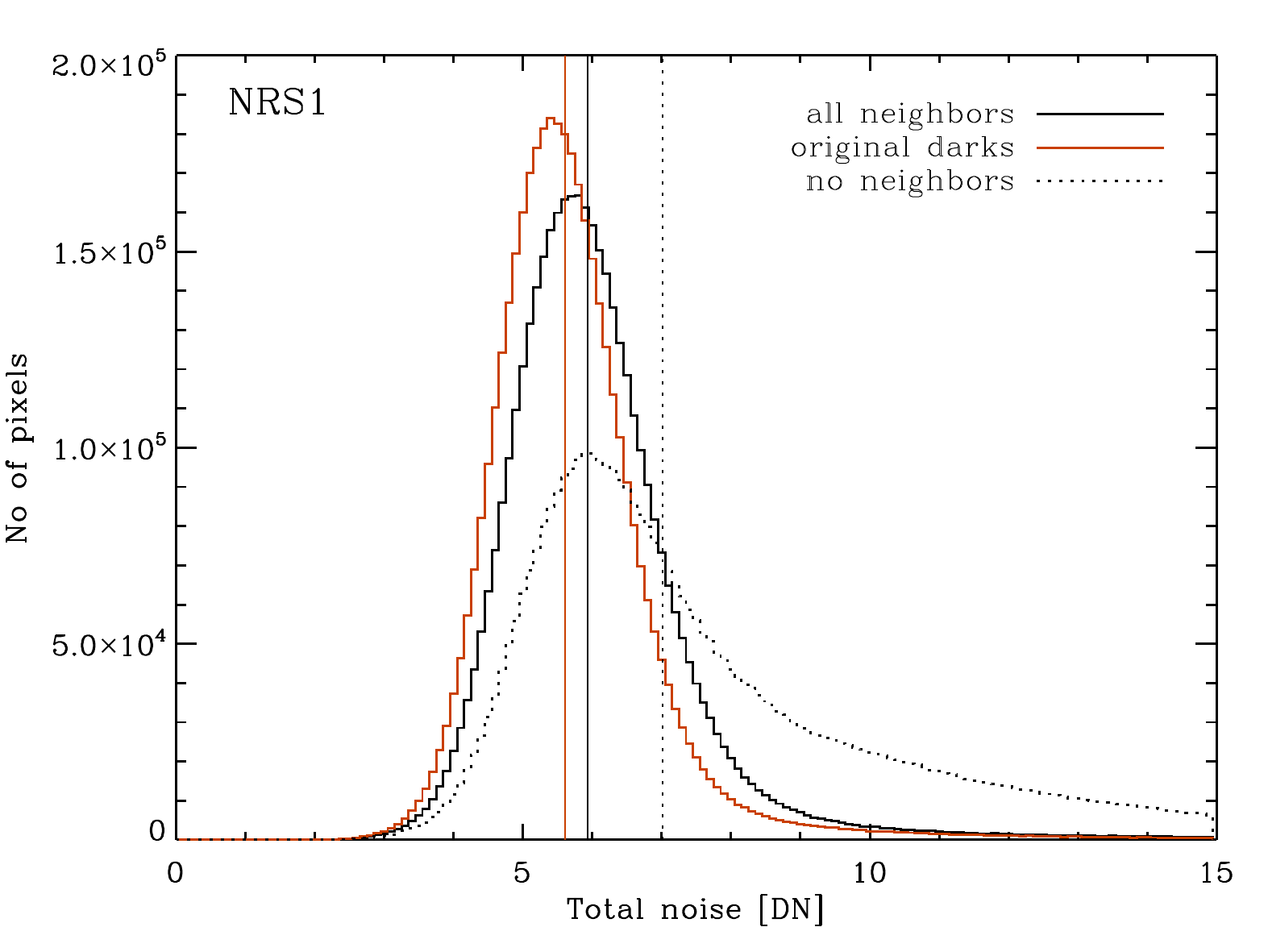}
\includegraphics[width=0.48\textwidth]{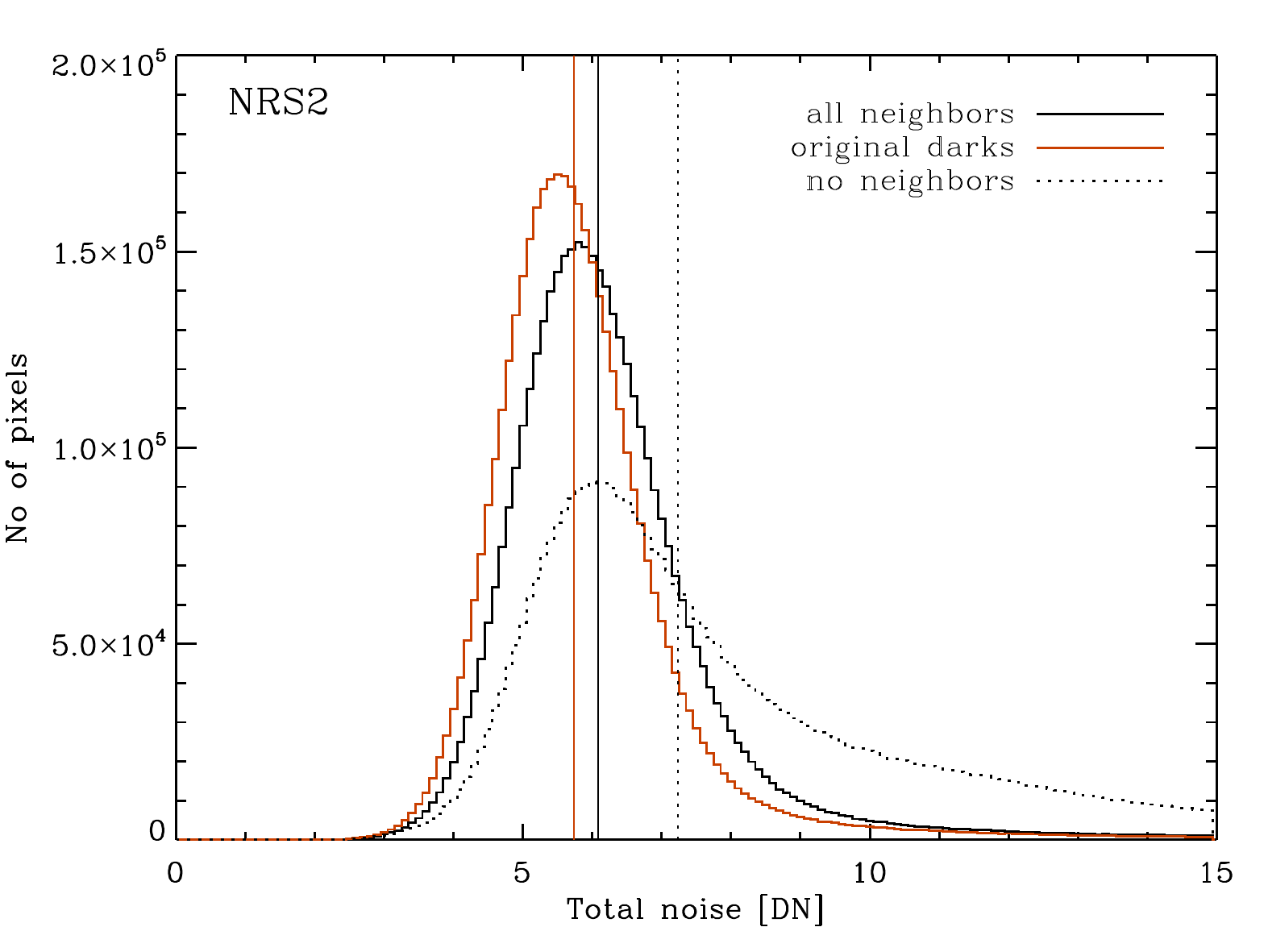}
\caption{\label{fig:noisehist}
Total noise histograms for NRS1 (left) and NRS2 (right). Red lines are for the original dark data with no added cosmic ray events, solid black lines for data with cosmic rays added and rejected with the 'all neighbors' method, and dotted black lines when not performing a second pass on the neighbors. In all cases, the histogram for the best rejection threshold $c_{th}$ is shown. The vertical lines denote the median total noise as presented in Table~\ref{tab:res}.}
\end{figure}

 Fig.\,\ref{fig:noisehist} shows the noise distributions for the
 original data and the data with injected CRs, each with and without
 second pass neighbor processing. The histograms demonstrate that a
 significant tail of pixels with higher noise remains in the processed
 darks, unless neighbors are also considered. This is reflected in the
 median noise values; compared to the original data without injected
 CRs, the total noise increases by $\sim$6\,\% for NRS1 and
 $\sim$7\,\% for NRS2, respectively, when a second pass on neighbors
 is performed, while increasing to $\sim$26\,\% (NRS1) and 28\,\%
 (NRS2) higher total noise with no second pass. The visual appearance
 of the same count-rate image processed with the different options for
 considering neighbors, shown in Fig.\,\ref{fig:vis} of the Appendix,
 also demonstrates that the `all-neighbors' choice delivers the
 cleanest results.

To study the effects of CRs on frame-averaged data, we used the same
traditional darks taken at OTIS, with and without injected cosmic ray
events, and frame-averaged them as would be done onboard. We repeated
the analysis described above, focusing on three test cases: {\em i )}
$c_{th} = [1.5\ldots 6.5]$ in steps of $0.2$ with no second pass on
neighbors (also for original data without simulated CRs); {\em ii)}
$c_{th} = [2.5\ldots 6.5]$ in steps of $0.2$ with cluster neighbor
detection enabled at $c_{nb} =0$; {\em iii)} $c_{th} = [2.5\ldots6.5]$ 
in steps of $0.2$ with all pixel neighbor detection enabled at
$c_{nb} = 0$, as long as the parent event causes a jump of at least 200
DN. The total noise results for the frame-averaged data is presented
in Figure~\ref{fig:noiseavg}, and the best value of $c_{th}$ for each case is given in table~\ref{tab:resavg} below.

\begin{figure}[t]
\setlength{\unitlength}{1\textheight}
\centering
\includegraphics[width=0.48\textwidth]{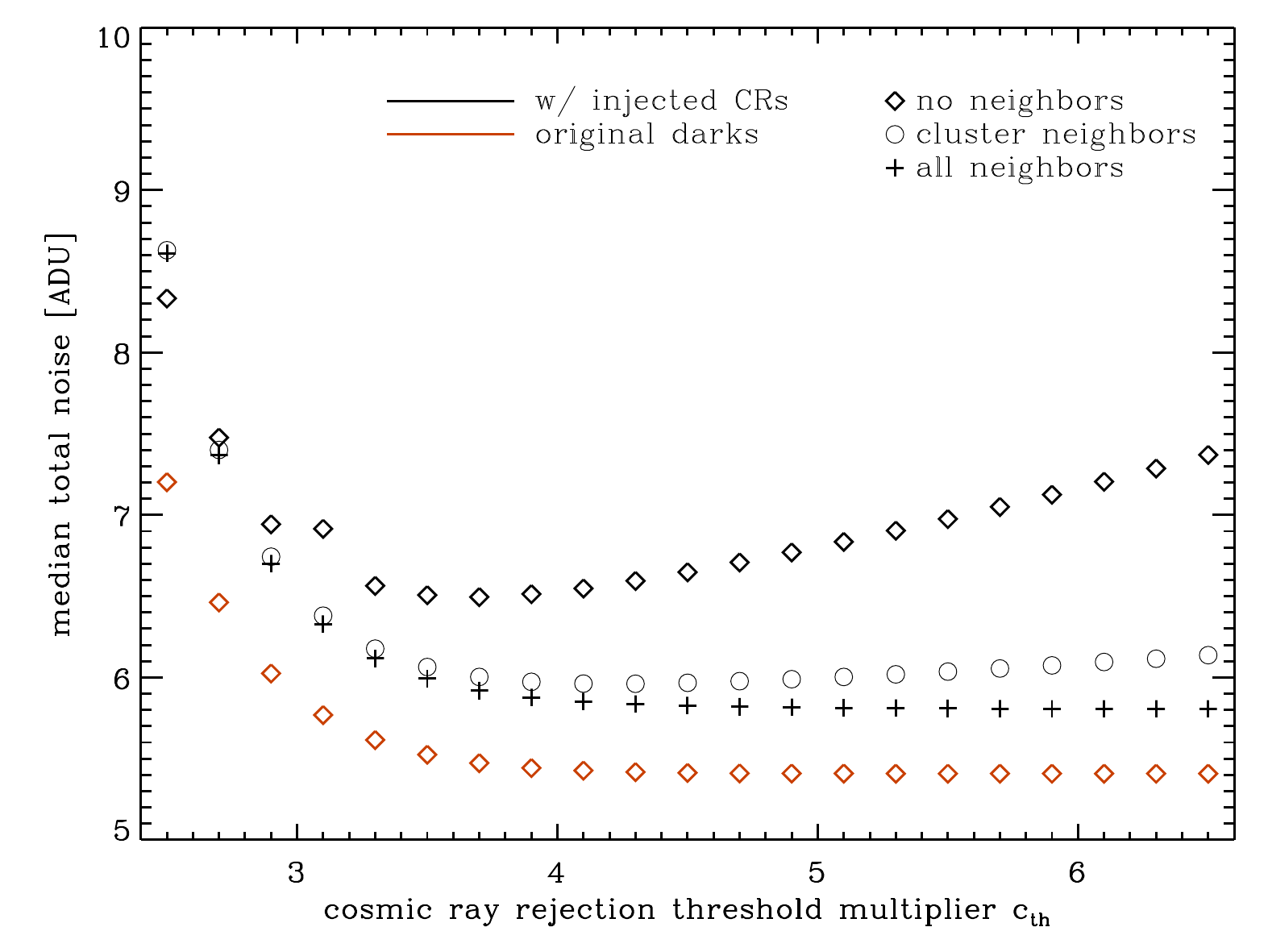}
\includegraphics[width=0.48\textwidth]{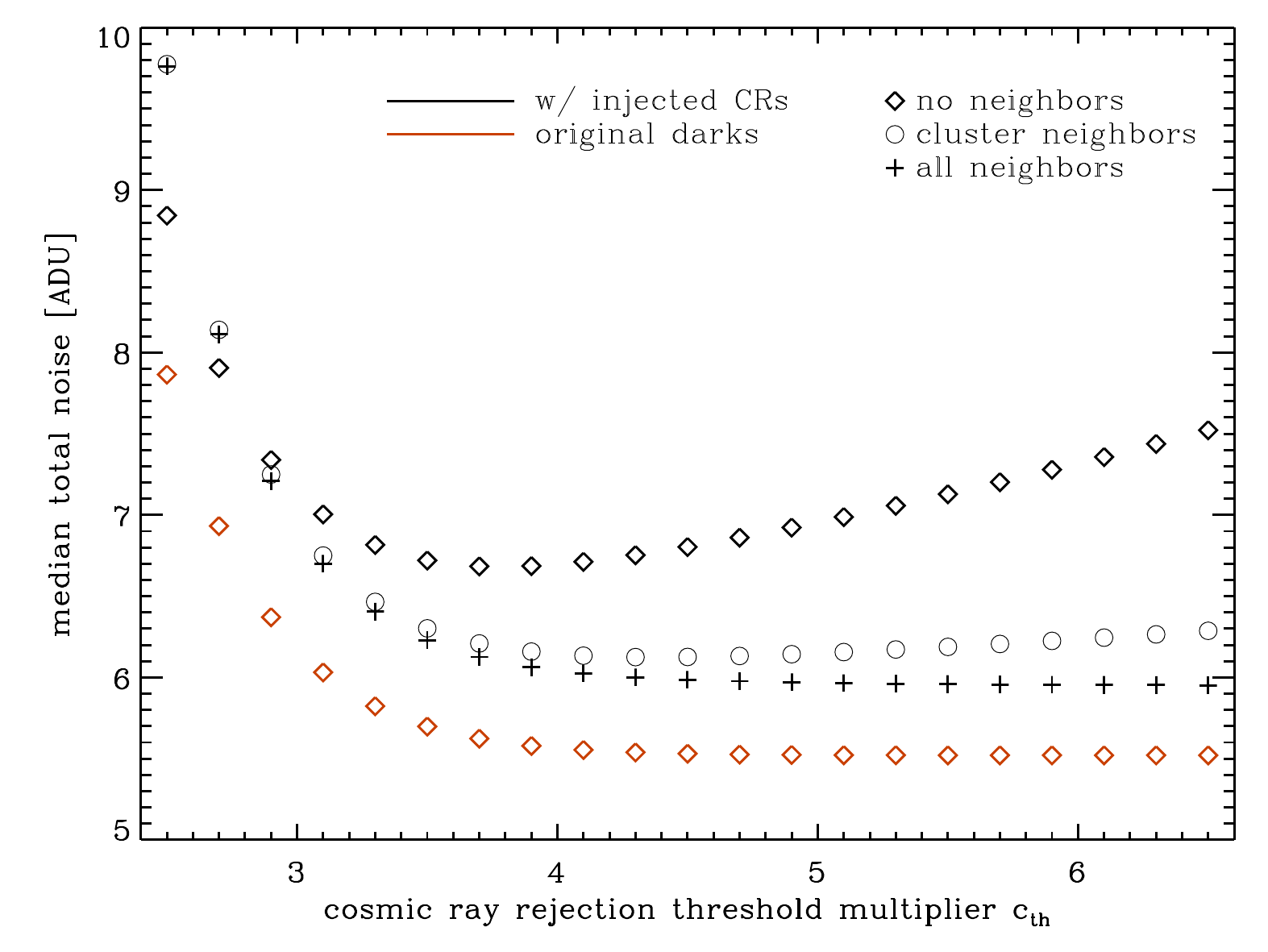}
\caption{\label{fig:noiseavg}Median total noise for NRS1 (left) and
  NRS2 (right) for the 25 frame-averaged dark exposures with (black
  symbols) and without (red symbols) simulated CR events.}

\end{figure}

\begin{table}[h]
\centering
\begin{tabular}{|l|r|r|r|r|}\hline
\multicolumn{1}{|c|}{$c_{nb}$}&\multicolumn{2}{c|}{NRS1}&\multicolumn{2}{c|}{NRS2}\\
&$c_{th}$&noise [e$^-$]&$c_{th}$&noise [e$^-$]\\\hline\hline
$\infty$ (no neighbors)&3.7&6.465&3.7&7.590\\
$0$ (all cluster neighbors)&4.3&5.933&4.3&6.956\\
$0$ (all neighbors)&5.9&5.779&6.5&6.759\\\hline
original data (no neighbors)&5.3&5.383&5.7&6.270\\\hline
\end{tabular}
\caption{\label{tab:resavg}Best parameters for CR detection threshold multiplier $c_{th}$ and the associated total noise for different values of $c_{nb}$ for the two NIRSpec detectors for frame-averaged data. The last row gives the total noise measured for the original (frame-averaged) dark data without simulated CRs added.}
\end{table}

The results are in line with those for the non-frame-averaged data, with a
second pass on all neighbors -- including those adjacent to severe
single pixel events -- being best. The optimal value of $c_{th}$ is
higher than for the non-frame-averaged case. For the best method with
all neighbors flagged, we recommend $c_{th} = 6.0$ for frame-averaged
data. The increase in total noise over the data without simulated cosmic
rays is $\sim$7\,\% for NRS1 and 8\,\% for NRS2, respectively, 
slightly larger than for the non-frame-averaged data. This can be
attributed to the 'dilution' of CR events in the
frame-averaged data, making detection more challenging, and the
fact that more data is lost due to a CR event when using
frame-averaging (longer effective groups, and one CR can lead to two
jumps in the ramp). Without a second pass on neighbors the noise
increase is $\sim$20\,\% (NRS1) and $\sim$21\,\% (NRS2).

Note that although the total noise values on the number of accumulated electrons (in e$^-$), for the frame averaged data, are slightly smaller than those for the non-averaged ones (cf. Table\,\ref{tab:res} and Table\,\ref{tab:resavg}), the total noise on the associated electron-rate (in e$^-/s$) are slightly larger, because the effective integration time, $t_{eff}$, 
decreases as $t_g$ increases with the number of averaged frames.

\subsection{IRS$^2$ readout} 
\label{sec:irs2}

Using the 30 dark exposures that were taken in IRS$^2$ readout mode, we processed both the original data and the data with cosmic rays added. The dark exposures taken in IRS$^2$ mode are longer than those in traditional readout mode (200 groups, corresponding to $\sim$3,000 seconds). Therefore, we processed both the full length exposures and only the first 65 groups of each integration, resulting in two data sets. The shorter integrations (65 groups, $\sim$1,000\,s) are comparable in length to the traditional data.

\begin{figure}[t]
\setlength{\unitlength}{1\textheight}
\centering
\includegraphics[width=0.48\textwidth]{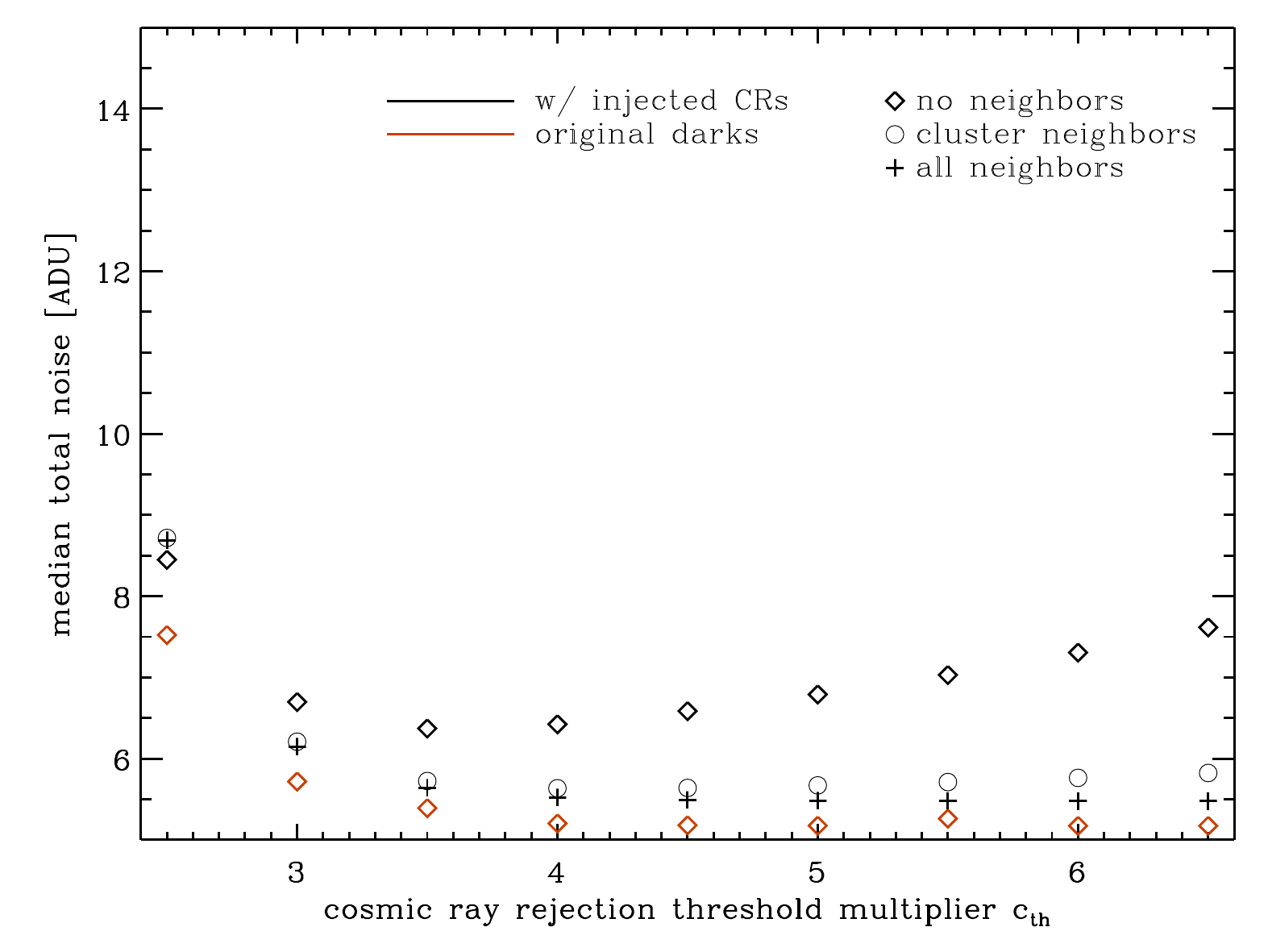}
\includegraphics[width=0.48\textwidth]{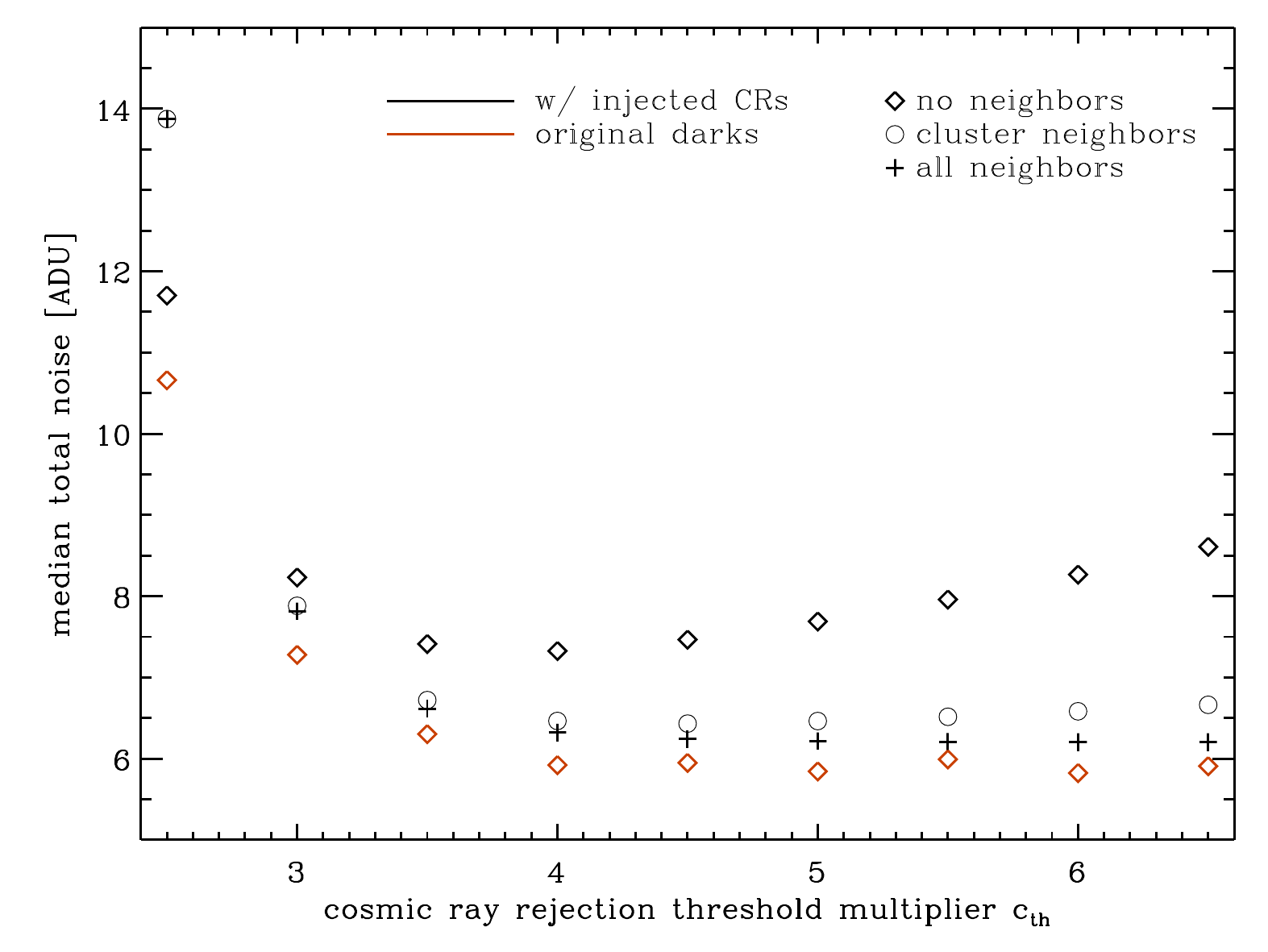}
\caption{\label{fig:noiseirs2short}Median total noise for NRS1 (left) and
  NRS2 (right) for the 30 dark exposures (NRSIRS2RAPID mode) with (black
  symbols) and without (red symbols) simulated CR events, processed up to 65 groups ($\sim$1,000 s integration time per exposure).}

\end{figure}

\begin{figure}[t]
\setlength{\unitlength}{1\textheight}
\centering
\includegraphics[width=0.48\textwidth]{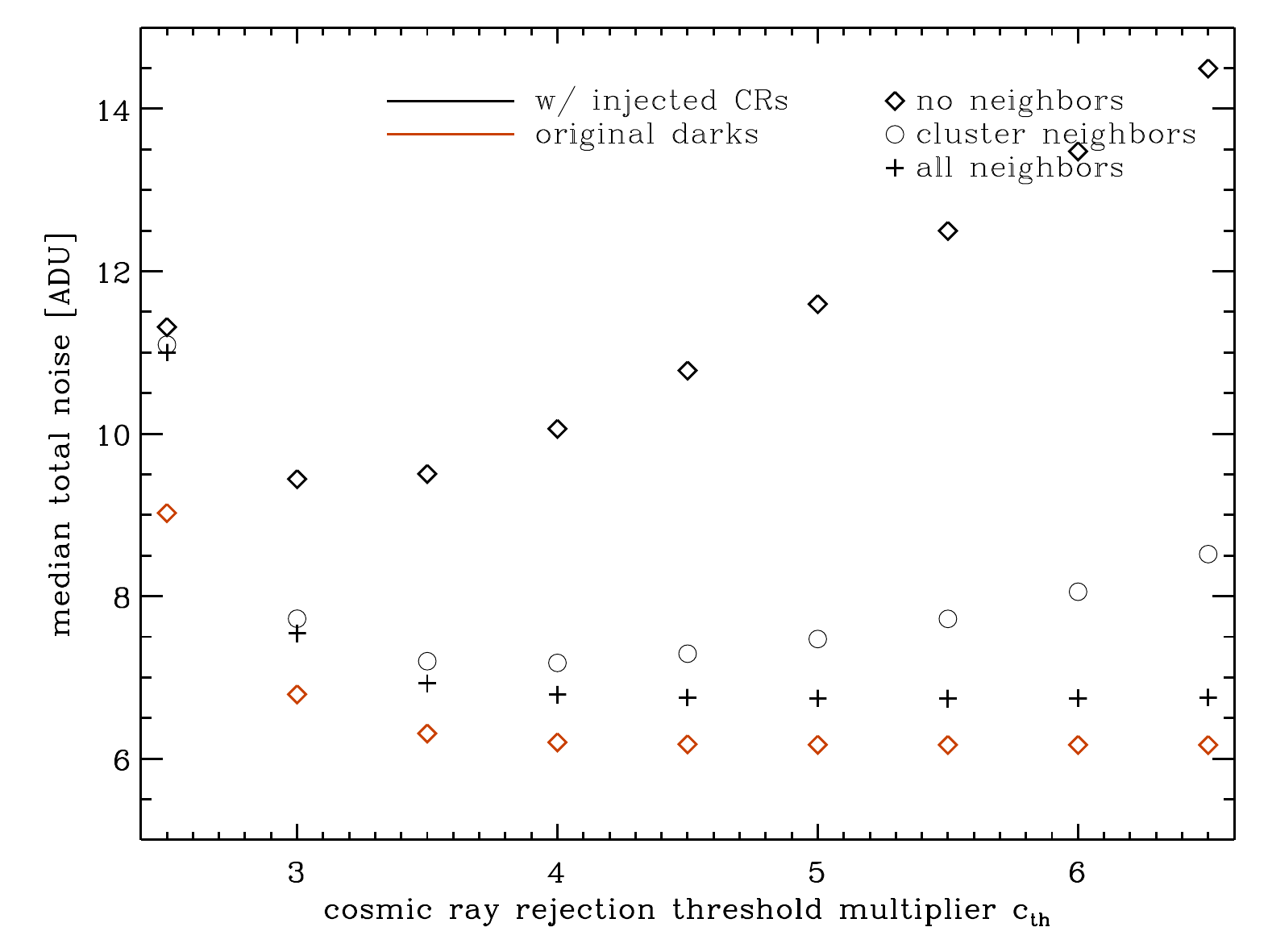}
\includegraphics[width=0.48\textwidth]{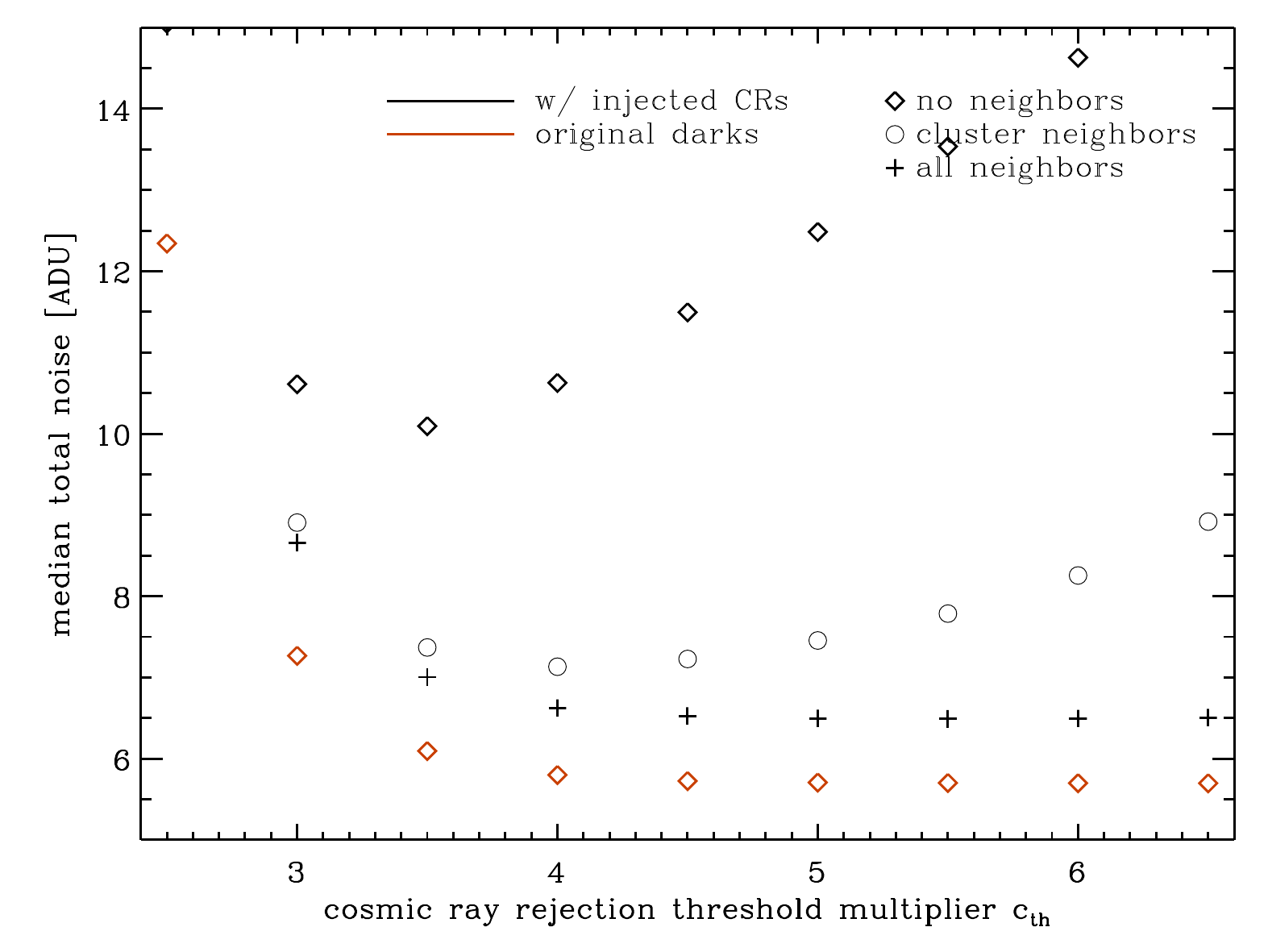}
\caption{\label{fig:noiseirs2}Median total noise for NRS1 (left) and
  NRS2 (right) for the 30 dark exposures with (black
  symbols) and without (red symbols) simulated CR events, for the full length integrations (200 groups, $\sim$3,000 s).}

\end{figure}

The processing of IRS$^2$ readout mode data is more time intensive, therefore we focused on three methods for the cosmic ray rejection - no neighbors, neighbors of clusters, and all neighbors - and used a coarser sampling of rejection threshold values $c_{th}$. The determination of total noise was achieved in the same way as outlined in section~\ref{sec:trad} for the traditional data. The results for both detectors are shown in Figures~\ref{fig:noiseirs2short} and \ref{fig:noiseirs2} for the short (65 groups) and full (200 groups) integrations, respectively.

The results are in line with those for the traditional data, with flagging of all neighbors yielding the lowest noise for CR-affected data. Table~\ref{tab:resirs2} lists the best values for each tested method, for both the short and full IRS$^2$ integrations.

\begin{table}[h]
\centering
\begin{tabular}{|l|r|r|r|r|r|r|r|r|}\hline
&\multicolumn{4}{|c|}{65 groups}&\multicolumn{4}{c|}{200 groups}\\
\multicolumn{1}{|c|}{$c_{nb}$}&\multicolumn{2}{c|}{NRS1}&\multicolumn{2}{c|}{NRS2}&\multicolumn{2}{c|}{NRS1}&\multicolumn{2}{c|}{NRS2}\\
&$c_{th}$&noise&$c_{th}$&noise&$c_{th}$&noise&$c_{th}$&noise\\\hline\hline
$\infty$ (no neighbors)&3.5&6.343&4.0&8.319&3.0&9.397&3.5&11.463\\
$0$ (all cluster neighbors)&4.0&5.610&4.5&7.317&4.0&7.146&4.0&8.099\\
$0$ (all neighbors)&4.5&5.452&6.0&7.045&5.5&6.709&5.5&7.369\\\hline
original data (no neighbors)&6.0&5.150&6.0&6.614&6.5&6.141&6.5&6.471\\\hline
\end{tabular}
\caption{\label{tab:resirs2}
Best parameters for CR detection threshold multiplier $c_{th}$
and the associated total noise (in e$^-$) for different values of $c_{nb}$ for the two NIRSpec detectors in IRS$^2$ readout mode. Results are for two different integration times of $\sim$1,000\,s (65 groups) and $\sim$3,000\,s (200 groups). The last row gives the total noise measured for the original dark data without simulated CRs added.}
\end{table}

For the short integrations, the increase in total noise with respect to the original data is comparable to the results for the traditional readout mode. For the best rejection method (two passes with all neighbors flagged), the increase in noise is $\sim$6\,\% for NRS1 and $\sim$6.5\,\% for NRS2, respectively. For the single-pass method (no neighbors), the noise increase is 23\,\% (NRS1) and 26\,\% (NRS2) compared to the original data. 

For the full (200 groups) integrations the increase in noise is significantly higher, with $\sim$9\,\% for NRS1 and $\sim$13\,\% for NRS2 when using the best (all neighbors) rejection. For the single-pass method, the increase is even more pronounced, at 53\,\% (NRS1) and 77\,\% (NRS2).

\section{Discussion and conclusion}
\label{sec:disc}

The aim of this work was to assess the impact of the CR bombardment expected at L2 on the noise performance of the (flight) NIRSpec detectors, and to investigate CR rejection algorithms based on the two-point difference method \citep{osf+1999, foh+00} 
to detect the CR-induced jump, and then split the signal ramp at the impacted frame to perform the ramp fitting on the separate segments. 

\cite{andgord2011} used simulations of JWST-like integration ramps to
study three methods to detect CR hits and concluded that the two-point
difference is the most effective in the shot-noise limited regime and
the best trade-off in terms of effectiveness and speed for read-noise
limited exposures. Additionally, the uncertainties associated with
this correction method are quasi-Gaussian, which simplifies the
process of choosing a rejection threshold.  Nevertheless, because of
the presence of significant inter-pixel signal coupling due to IPC in
these detectors and because the CR signature often involves more than
one pixel, applying this approach only to the pixel(s) where the hit
is initially detected leads to considerably worse total noise for our
detectors compared with ground-data: up to $\sim$ 30\% of the noise
level for integration lengths of 1,000 s and more than $\sim$ 75\% for
integrations of 3,000 s.  

Our analysis shows that significantly better
noise performance can be achieved with a second pass which addresses
the direct neighbors of CR-affected pixels and applies the same
ramp-splitting to them. This is in line with the results by
\cite{fap2006} who also found that interpixel correlation
substantially increases the footprint of CR impacts and treated  neighboring pixels
with a second pass in their pipeline for NICMOS data
reduction.  By including neighboring pixels in our CR mitigation
approach, we achieve noise levels that are only slightly elevated
compared to ground performances: by about 7\% for 1,000 s integrations
and 15\% for 3,000 s.

The increase in noise is mainly due to two effects. Some CRs will
cause saturation of the pixel so the slope can only be estimated from
data before the hit occurred, yielding a reduction in effective
integration time. Even if the cosmic ray event does not result in
saturation, the ramp needs to be split. While the full data loss is
low (typically one group time, up to two group times for
frame-averaged data per event), the two ramp segments need to be fit
individually, and the variance of the combined slopes is higher than
that of a single, uninterrupted ramp.  The detrimental effect of
cosmic rays is more pronounced for longer integrations, because more
data can be lost to early saturation due to a CR hit, and also more
pixels are affected by a CR, i.e. have their integration broken into
two or more segments.

Besides the impact on the noise statistics of the detectors presented
here, CRs can also lead to systematic effects and bias the
measurements in more subtle ways that are not addressed by our
analysis. In first instance, the choice of threshold value impacts the
fraction of CR-affected pixels that will go undetected, and these will
inevitably bias the overall signal distribution to a higher value.
For each of the CR affected pixels, in every exposure, we computed the
difference in count rate between the original data and the CR-injected
data processed with a range of $c_{th}$ values (all-neighbors method);
the mean excess signal (of all CR-affected pixels, excluding hot
pixels) increases approximately linearly with the threshold, for
$c_{th}$ in range 2.5$-$6.5. This bias (i.e. offset of the mean from
zero) is, however, small compared to the count-rate variations, less
than 10\% of the RMS of the difference distribution, for $c_{th} =
6.5$.

Near-IR detectors exhibit the effect of persistence
\citep{regberg2018}. The physical process that causes the persistence
has been ascribed to charge 'traps', i.e. defects in the semiconductor
material that capture charge during the stimulus image and then slowly
release it over time. Therefore the persistent signal from strong CR
hits, corresponding to a count of more than $\sim$ 50,000 electrons,
can pollute subsequent images as well as affect the linearity of the
pixel response after the hit, as some of the charge-traps ``leak"
additional spurious signal.  Furthermore, following such a strong CR
hit, the pixel will be operating at a higher count level, closer to
its inherent non-linear regime, where the non-linearity correction is
applied by the ramp-to-slop pipeline. This may also impact somewhat
the total noise figure, as the accuracy of the flux estimate is
affected by the uncertainties of the linearity correction.  With the
CR fluence assumed in this work, in a 1,000 second integration, fewer
than 0.3\% of the pixels (about 10,000) are affected by an event that
introduces more than 50,000 electrons. For specialized programs aiming
to minimize potential systematic errors, one option may simply be to
flag and discard disturbed pixels; in this context, the method
presented here allows a more comprehensive flagging of affected
pixels.

We also note that, in our simulations, CRs were only injected into the light sensitive pixels (central 2040 x 2040 region of the detector) and not
the four pixel wide 'frame' of reference pixels (or the interleaved reference pixels in case of the IRS$^2$ readout data). Indeed, the simulated events are modeled for charge deposited into an
active layer of HgCdTe, and not for effects on the underlying read out integrated
circuit (ROIC). This is justified by detector data from the Hubble Space Telescope Wide Field Camera~3 which is equipped with a previous generation of Teledyne near-IR detectors (HAWAII-1R). Those data indicate that CRs do not appear to affect significantly the reference pixels (Ben Sunnquist, personal communication). 

In conclusion, our analysis has allowed us to develop an effective way to deal with the CR bombardment expected for NIRSpec detectors in flight, and to show that with a tailored rejection algorithm, the CR hits will cause only a small deterioration of the detectors' total noise level, with an increase of $\sim$ 7\%, for the bench mark 1,000 s exposures. This corresponds to a 7\% decrease in the limiting sensitivity 
of the instrument for the medium and high-spectral resolution modes ($R=1000$ and $R=2700$) that, for faint targets, are fully detector-noise limited; for the low-spectral resolution configurations ($R=100$) that are not typically detector noise limited (because background noise is a significant contributor), the impact of the noise increase is lower, 3-4\% depending on wavelength. Such small decreases in sensitivity are not a concern for the achievement of the NIRSpec scientific goals. 

\begin{acknowledgments}
G. Giardino thanks Ralf Kohley for useful discussion.
\end{acknowledgments}

\appendix
\section{Cosmic Ray Detection Efficiency}
\label{sec:efficiency}

\begin{figure}
\centering
\includegraphics[width=0.4\textwidth]{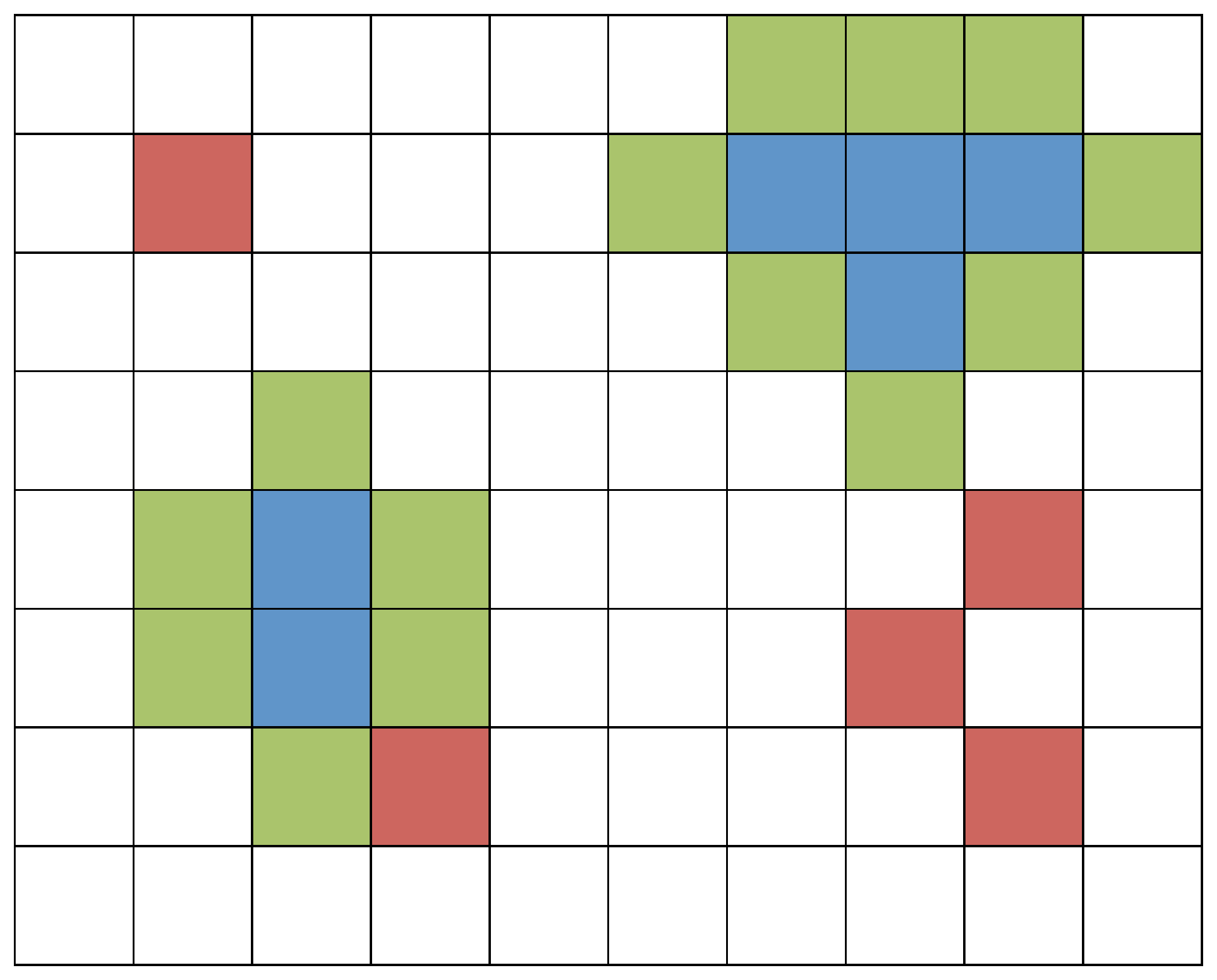}
\includegraphics[width=0.4\textwidth]{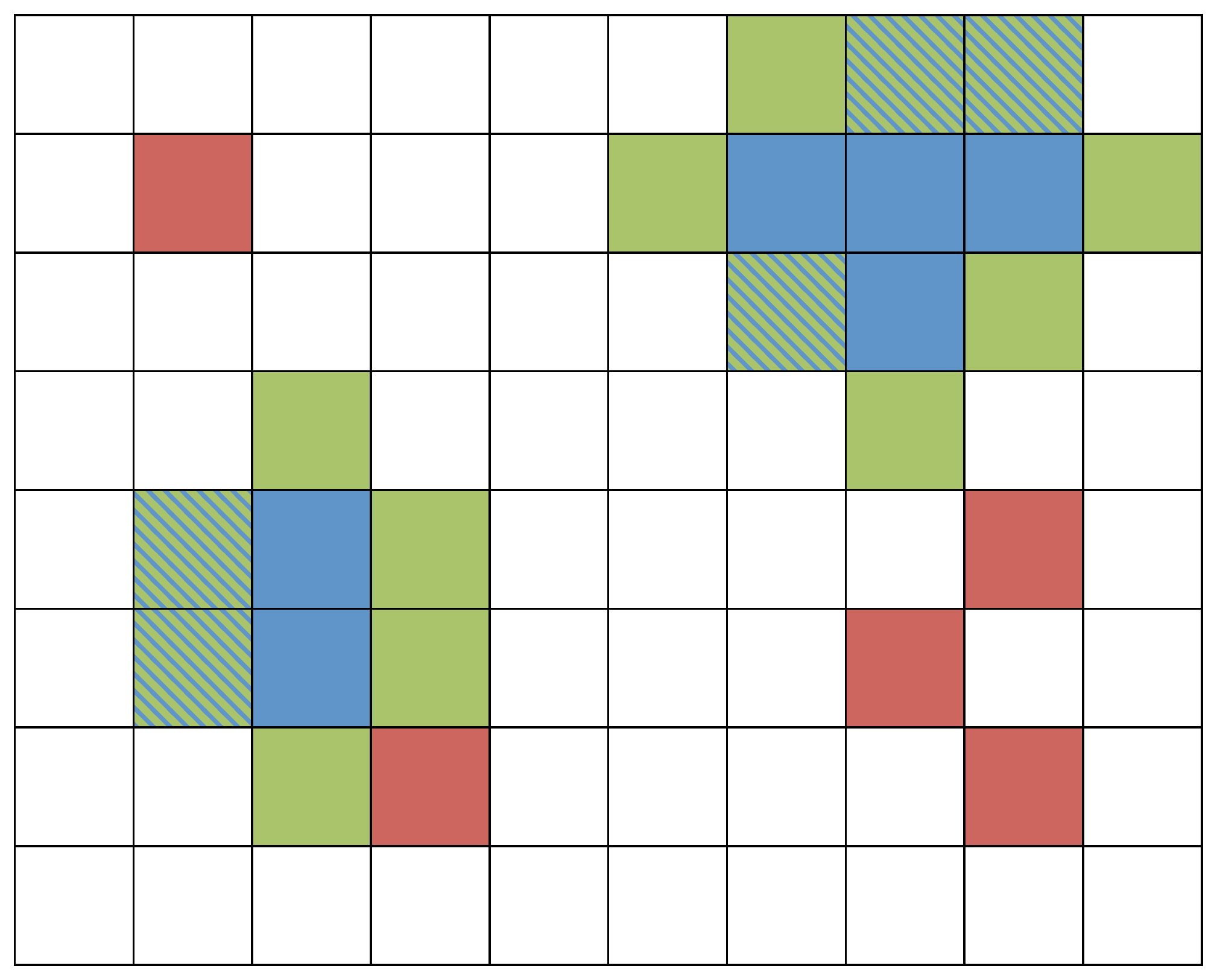}
\caption{\label{fig:clusters}Schematic of a grid of 10 by 8 pixels (white squares) with some cosmic hits / outliers detected using a threshold $c_{th}$ (left). Outliers can occur isolated (red) or in clusters (blue). The direct neighbors of cluster pixels are marked in green. These pixels can then be given more scrutiny in a second detection pass that uses a lower detection threshold $c_{nb}$. In this example five additional pixels were found with this lower threshold, marked with blue/green stripes (right).}
\end{figure}

As described in Sect.\,\ref{sec:algo} our pipeline offers the
possibility to perform a second pass of outliers detection for pixels
at a lower threshold for the neighbors of pixels that had a detected
CR in the first pass, with two options for selecting
neighbors: {\em i)} directly adjacent pixels of any pixel that has a
CR event of at least a certain magnitude, or {\em ii)}
neighbors have to be adjacent to a cluster of pixels with detections
at the same group. A cluster consists of two or more pixels that are
directly adjacent, i.e. neighbor at the top, bottom, left or right, as
shown in Fig.\,\ref{fig:clusters}.

\begin{figure}
\setlength{\unitlength}{1\textheight}
\centering
\includegraphics[width=0.48\textwidth]{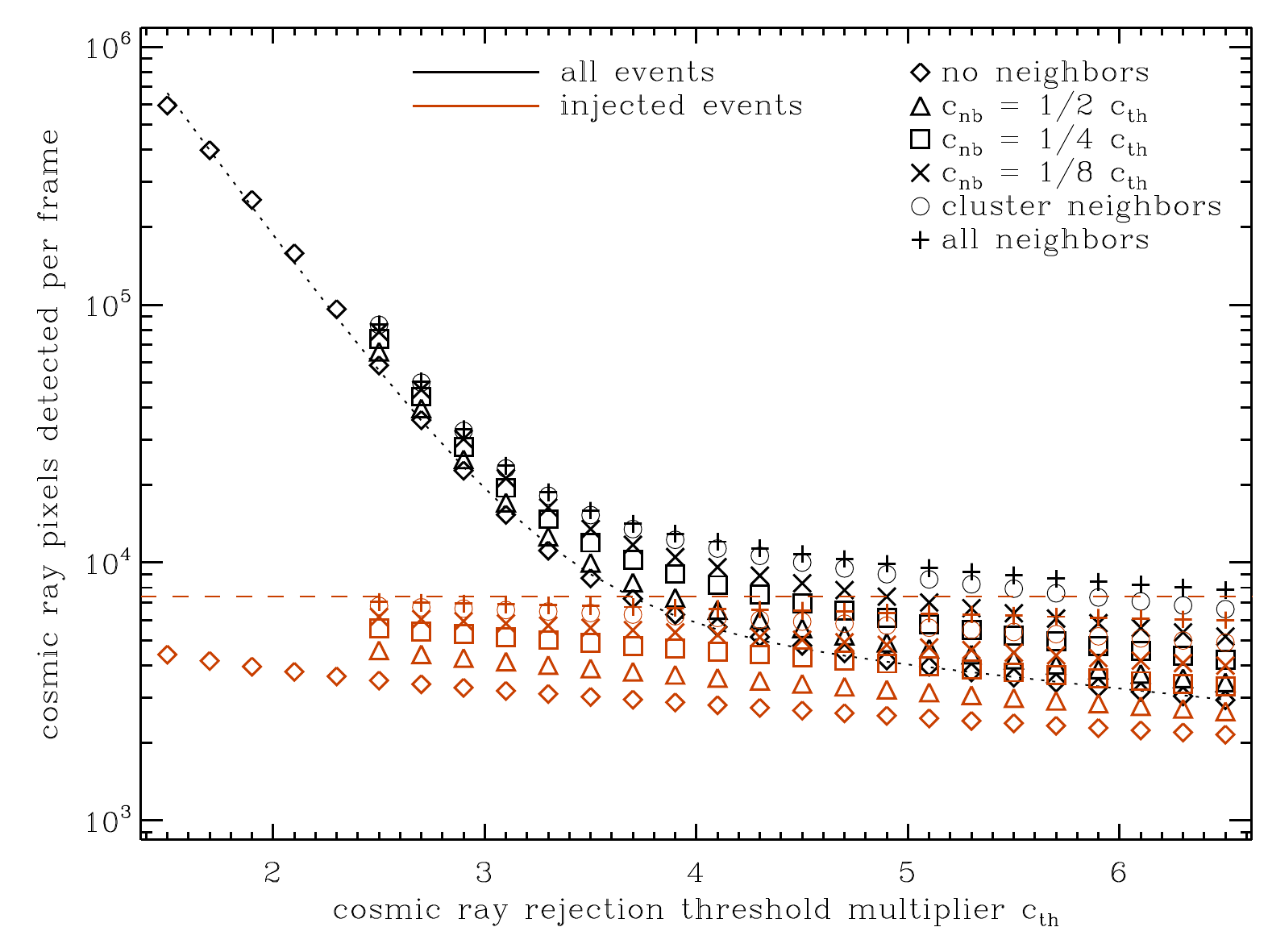}
\includegraphics[width=0.48\textwidth]{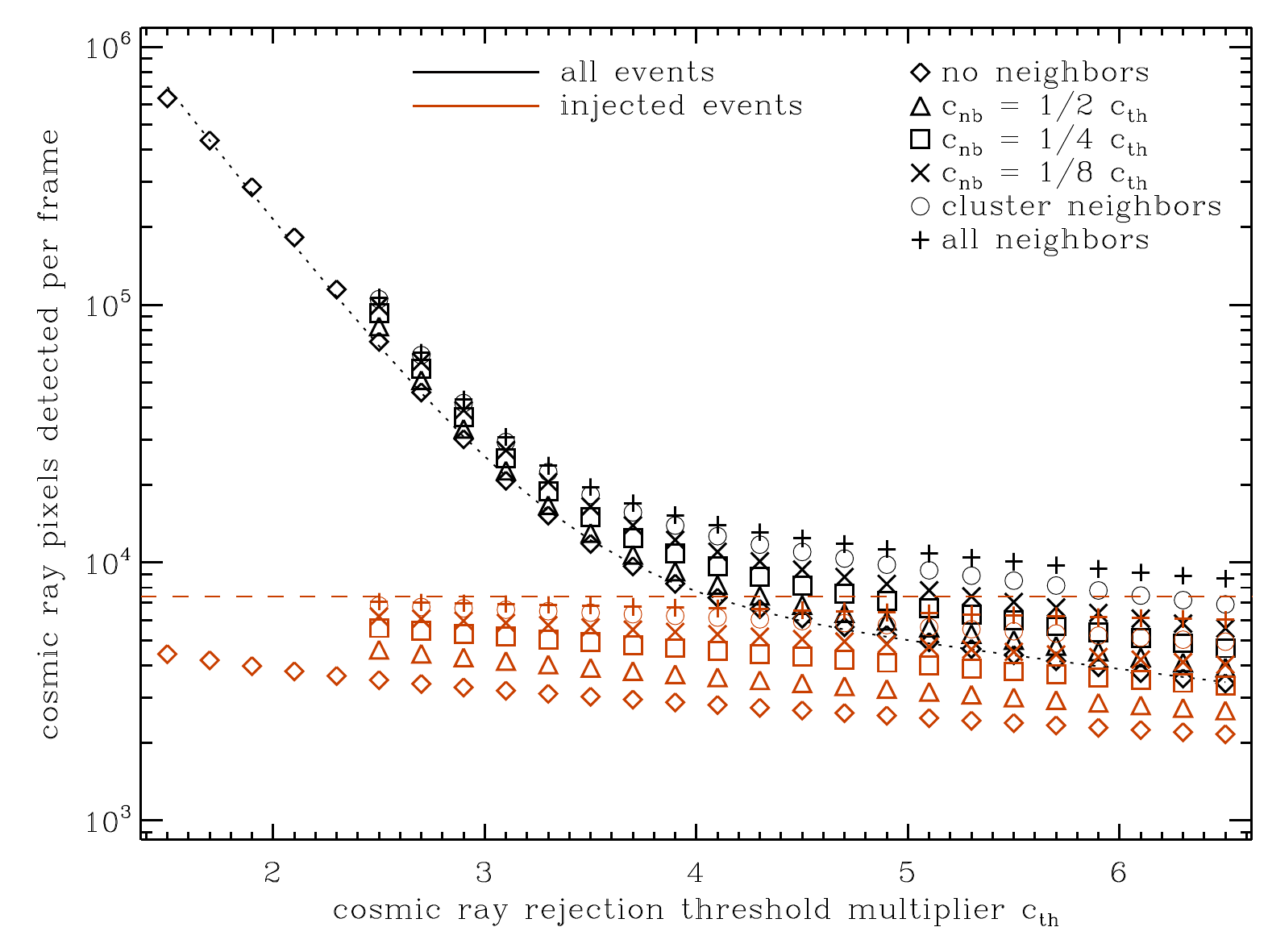}
\caption{\label{fig:detections}Number of pixels with detected outliers
  per frame as a function of detection thresholds $c_{th}$ and
  $c_{nb}$ for NRS1 (left) and NRS2 (right). The black
  symbols denote all detections, the red symbols are for detected
  events that were injected. The dashed red line shows the total
  number of injected events. The dotted black line denotes a fit to
  the data with no neighbor detection according to
  equation~\ref{eq:fit}.}
\end{figure}

To asses the efficiency of CR detections by the pre-processing pipeline, the detected events were recorded
and compared to the injected simulated events that are detectable,
i.e.\ occur after the first group is read and do not lead to
saturation. The number of pixels with detected events per frame for
both detectors as a function of detection thresholds $c_{th}$ and
$c_{nb}$ is shown in Figure~\ref{fig:detections}.

As expected, lower values of $c_{th}$ and $c_{nb}$ yield higher
detection rates. In general, the total number of detected events seems
to follow a two component exponential function with a knee at $c_{th}
\sim 4$, rapidly increasing with low values of $c_{th}$. For the case
with no neighbors (equivalent to $c_{nb} \ge c_{th}$) the number of
pixels per frame with detections is well fitted by the following
function:
\begin{equation}\label{eq:fit}
n_{detections} = \alpha_1 \times 10^{\beta_1 c_{th}} + \alpha_2 \times 10^{\beta_2 c_{th}},
\end{equation}
with the parameters presented in table~\ref{tab:fit} below.
\begin{table}[!h]
\centering
\begin{tabular}{|l|r|r|r|r|}\hline
Detector&$\alpha_1$&$\beta_1$&$\alpha_2$&$\beta_2$\\\hline\hline
NRS1&$3.17\times10^7$ &$-1.124$& $1.06\times10^4$&$ -0.086$\\
NRS2&$2.64\times10^7$ &$-1.055$& $1.55\times10^4$&$ -0.101$\\\hline
\end{tabular}
\caption{\label{tab:fit}Fit parameters for both detectors for the number of detected pixels with CRs in the case of no neighbor detection.}
\end{table}

Detection numbers for other values of $c_{nb}$ can also be fitted by the function in equation~\ref{eq:fit}, but are not shown here.

\section{Visual Appearance} 
\label{sec:appearance}

\begin{figure}
\setlength{\unitlength}{0.1\textwidth}
\begin{picture}(10,10)
\put(0,5.1){\includegraphics[width=0.49\textwidth]{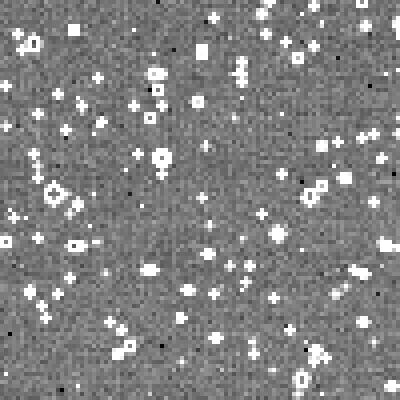}}
\put(5.1,5.1){\includegraphics[width=0.49\textwidth]{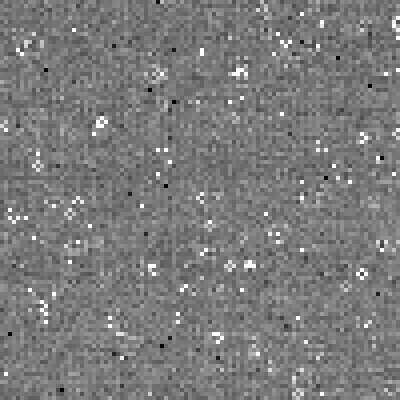}}
\put(0,0){\includegraphics[width=0.49\textwidth]{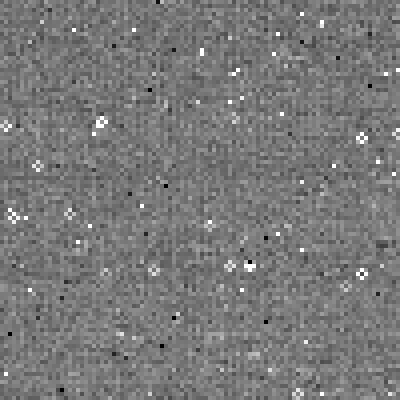}}
\put(5.1,0){\includegraphics[width=0.49\textwidth]{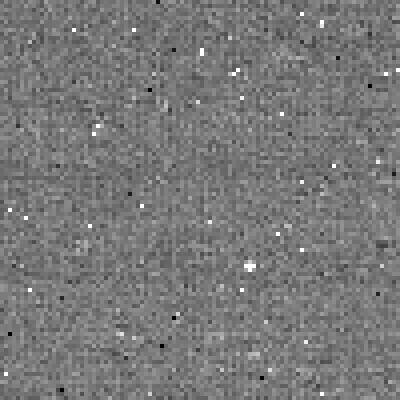}}
\put(0.1,5.2){\large \color{white}{no CR rejection}}
\put(5.2,5.2){\large \color{white}{\Large c\footnotesize{th }\large = 3.5 ; no neighbors}}
\put(0.1,0.1){\large \color{white}{\Large c\footnotesize{th }\large = 3.9 ; all cluster neighbors}}
\put(5.2,0.1){\large \color{white}{\Large c\footnotesize{th }\large = 4.9 ; all neighbors}}
\end{picture}

\caption{\label{fig:vis}Count rate image of a 100$\times$100 pixel
  region with injected CRs on NRS1 dark exposure (traditional
  readout). Scale is black to white from -0.1 to 0.1 DN/s. Note that
  the great majority of residual bright pixels in the bottom right
  panel are not due to undetected CRs but to the higher shot noise of
  hot-pixels; as can be seen in the image, they are matched by a
  similar number of black pixels, which are also hot-pixels (our
  images are dark-subtracted)}
\end{figure}

In Figure~\ref{fig:vis} we show a region of the count rate maps of an NRS1 dark exposure in traditional readout mode with injected cosmic ray events with different cosmic ray rejection methods and thresholds:
\begin{enumerate}
\item no cosmic ray rejection
\item cosmic ray rejection with $c_{th} = 3.5$, but no neighbors flagged
\item cosmic ray rejection with $c_{th} = 3.9$ and all cluster neighbors flagged ($c_{nb} = 0$)
\item cosmic ray rejection with $c_{th} = 4.9$ and all neighbors flagged ($c_{nb} = 0$), including neighbors of single pixels with a severe ($\ge$ 200 DN) jump in their parent
\end{enumerate}

The value of $c_{th}$ was chosen to yield the best total noise for the given detection method, as presented in table~\ref{tab:res}.


Without cosmic ray rejection, events are clearly visible and manifest as elevated count rates in usually five or more pixels. The cross-like shape of some events is due to the IPC. The centers of some cosmic ray events seem to have low count rates consistent with pixels not hit. This is due to saturation by the simulated event, so that the pixel is then flagged as saturated from that group on and the affected data is not used to determine the slope. Therefore, such events are filtered out even without jump detection.
 
Cosmic ray rejection without neighbor detection removes a significant fraction of cosmic rays, but tends to leave ring like structures around stronger events.
 
Cosmic ray rejection with flagging all cluster neighbors gives an even better visual appearance in terms of rejected cosmic rays. The leftovers are mostly the four outer pixels of five-pixel (cross-like) events. This is because, for those events, only a single pixel is identified as a CR in the first pass. Therefore, it does not qualify as belonging to a cluster and thus no neighbors are flagged (see also Figure~\ref{fig:clusters}).

Flagging neighbors to a single pixel, with an event (jump) of at least 200 DN, as well as cluster neighbors, gives the best visual appearance (and total noise, as shown in section~\ref{sec:res} above).



\end{document}